\documentclass[floatfix, prl, superscriptaddress, onecolumn, notitlepage, 11pt]{revtex4-1}

\usepackage[dvips]{graphicx}	% Include figure files
\usepackage{dcolumn}% Align table columns on decimal point. Column format d. 
\usepackage{bm}					% bold math
\usepackage{amsmath,amsbsy,amsfonts,revsymb}

\newcommand{\Vcoeff}[3]{\ensuremath{{{}^{#3}\!V_{#1}^{#2}}}}
\newcommand{\Acoeff}[3]{\ensuremath{{{}^{#3}A_{#1}^{#2}}}}
\newcommand{\Ffun}[2]{\ensuremath{F_{#1}^{#2}}}

\newcommand{\SpherHar}[1]{\ensuremath{Y_{#1}}}

\newcommand{\Clebsch}[2]{\ensuremath{\mathcal{C}_{#1}^{#2}}}

\def\Htwoatsixty/{H$_2$@C$_{60}$}
\def\Dtwoatsixty/{D$_2$@C$_{60}$}
\def\HDatsixty/{HD@C$_{60}$}
\def\Htwoatseventy/{H$_2$@C$_{70}$}
\def\Dtwoatseventy/{D$_2$@C$_{70}$}
\def\TwoHtwoatseventy/{(H$_2$)$_2$@C$_{70}$}
\def\Csixty/{C$_{60}$}
\def\Cseventy/{C$_{70}$}
\def\Htwo/{H$_2$}
\def\para/{{\it para\/}}
\def\ortho/{{\it ortho\/}}
\def\Dtwo/{D$_2$}
\def\Wateratsixty/{H$_2$O@C$_{60}$}
\def\Water/{H$_2$O}
\def\wn/{\,cm$^{-1}$}
\def\area/{\,cm$^{-2}$}

\def\svec/{\mathbf{s}}

\def\rivec/{\mathbf{r}_i}
\def\ri/{r_i}
\def\thetai/{\theta_i}
\def\phii/{\phi_i}
\def\riunitvec/{\hat{\mathbf r}_i}
\def\thetaiunitvec/{\hat{\pmb\theta}_i}
\def\phiiunitvec/{\hat{\pmb\phi}_i}

\def\angulari/{\Omega_{i}}

\def\RCMvec/{\mathbf{R}}
\def\RCM/{R}
\def\RCvec/{\mathbf{R}_G}
\def\thetaCM/{\Theta}
\def\phiCM/{\Phi}
\def\angularCM/{\Omega}

\def\CMshift/{\mathbf{x}}

\def\RCivec/{\mathbf{R}_i}
\def\RCi/{R_i}
\def\thetaCi/{\Theta_i}
\def\phiCi/{\Phi_i}
\def\RCiunitvec/{\hat{\mathbf R}_i}
\def\thetaCiunitvec/{\hat{\pmb\Theta}_i}
\def\phiCiunitvec/{\hat{\pmb\Phi}_i}
\def\angularCi/{\Omega_{Ci}}

\def\angularCi/{\Omega_{Ci}}

\def\angularHtwo/{\Omega_s}

\def\psiTv{\psi^{v}}
\def\PsiTv{\Psi^{v}}

\def\psiV{\psi^{vib}}

\def\MJ{M_J}

\def\ML{M_L}
\def\MLambda{M_\Lambda}

\def\MLambdai{M_{\Lambda_i}}
\def\mlambda{m_\lambda}

\def\H{\mathcal H}
\def\HVR{\H^{\mathrm{vib - rot}}}
\def\HVRv{{}^v\HVR}
\def\EVR{E^{\mathrm{vib - rot}}}
\def\EVRvJ{{}^v\!\EVR_J}
\def\EVRvJz{{}^v\!\EVR_{JJ_z}}

\def\ket#1{\left| #1 \right>}
\def\bra#1{\left< #1 \right|}
\def\Vv{{}^v V}
\def\Vvzero{\Vv^0}
\def\Vvprime{\Vv^\prime}
\def\wzeroV{\omega_0}
\def\wzeroT{\omega_0^T}
\newcommand{\wzeroTv}[1]{^#1\!\wzeroT}
\def\wzeroTz{\omega_z^T}
\def\wzeroTxy{\omega_{xy}^T}
\def\vkappa{^v\!\!\kappa}

\def\Inv/{\hat{I}}
\def\Unity/{\hat{E}}

\begin{document}

\title{Infrared spectroscopy of small-molecule endofullerenes}

\author{T. R{\~o}{\~o}m}
\email{toomas.room@kbfi.ee}
\author{L. Peedu}%C70
\author{Min Ge}
\author{D. H{\"u}vonen}
\author{U. Nagel}
\affiliation{National Institute of Chemical Physics and Biophysics, Akadeemia tee 23, 12618 Tallinn, Estonia}
\author{Shufeng Ye}
\author{Minzhong Xu}
\author{Z.  Ba\v{c}i\'{c}}
\affiliation{Department of Chemistry, New York University, New York, New York 10003, USA} 
\author{S. Mamone}
\author{M. H. Levitt}
\author{M. Carravetta}
\affiliation{School of Chemistry, Southampton University, Southampton SO17 1BJ, United Kingdom}
\author{J. Y.-C. Chen}%op conversion
\author{Xuegong Lei}
%\author{Yongjun Li}%Water
\author{N. J. Turro}
\affiliation{Department of Chemistry, Columbia University, New York, New York 10027,USA}
%\author{K. Kurotobi}%Water
\author{Y. Murata}
\author{K. Komatsu}
\affiliation{Institute for Chemical Research, Kyoto University, Kyoto 611-0011, Japan}

\date{\today}

\begin{abstract}
Hydrogen is one of the few molecules which  has been incarcerated in the molecular cage of \Csixty/ and forms endohedral supramolecular complex  \Htwoatsixty/.
In this confinement hydrogen acquires new properties.
Its translational motion becomes quantized and is correlated with its rotations.
We applied infrared spectroscopy to study the dynamics of hydrogen isotopologs \Htwo/, \Dtwo/ and HD incarcerated in \Csixty/.  
The translational and rotational modes appear as side bands to the hydrogen vibrational mode in the mid infrared part of the absorption spectrum.
Because of the large mass difference of hydrogen and \Csixty/ and the high symmetry of \Csixty/ the problem is identical to a problem of a vibrating rotor moving in a three-dimensional spherical potential.
The translational motion within the \Csixty/ cavity  breaks the inversion symmetry and induces optical activity of \Htwo/.
We derive potential, rotational, vibrational and dipole moment parameters from the analysis of the infrared absorption spectra.
Our results were used to derive the parameters of a pairwise additive five-dimensional potential energy  surface for \Htwoatsixty/.
The  same parameters were used to predict \Htwo/ energies inside \Cseventy/[Xu et al., J. Chem. Phys., {\bf 130}, 224306 (2009)]. 
We compare the predicted energies and the low temperature infrared absorption spectra of \Htwoatseventy/.  
\end{abstract}

%\pacs{33.20.Vq, 36.20.Ng, 63.22.+m, 78.30.Na}
\keywords{endohedral, fullerene, para ortho hydrogen, water, infrared}

\maketitle
%\newpage

%=============================
\section{Introduction}
%=============================

A small cavity inside the fullerene cage is a potential trapping site of atoms and has attracted attention of scientists from the moment of discovery of \Csixty/ \cite{Kroto1985}.
The demonstration of formation of La@\Csixty/ after laser bombardment of La-impregnated graphite was immediate \cite{Heath1985}.
Since then the field of studies of endohedral fullerenes has expanded.
Endohedral fullerenes with noble gas (He and Ne in \cite{Saunders1993}, Ar, Kr, Xe \cite{Saunders1994}),  nitrogen \cite{Mauser1997} or phosphorus \cite{Larsson2002} atoms and with metal clusters \cite{Dunsch2007Review}  are made under extreme conditions using arc discharge, ion bombardment or high pressure and high temperature treatment.

Extreme methods are not suitable for encapsulation of small molecules.
A different approach, ``chemical surgery'', was applied by Rubin when he  made first open-cage fullerene \cite{Rubin1997} with a orifice large enough to load it with  $^3$He or \Htwo/ using  less extreme temperature and pressure \cite{Rubin2001}.
Soon Y. Murata, M. Murata and K. Komatsu  synthesized another opened-cage derivative of \Csixty/  and achieved 100\% yield in filling with \Htwo/ \cite{Murata2003JACS}. 
There a generation of closed-cage \Htwoatsixty/ was observed in the process of matrix-assisted laser desorption/ionization time-of-flight  mass spectrometry analysis of this opened-cage complex.
Chemical methods were developed to close opened-cage fullerenes and \Htwoatsixty/ was produced   in  milligram quantities  \cite{Komatsu2005,Murata2006}.
To accommodate two hydrogen molecules a  cavity larger than \Csixty/  is needed.
Two \Htwo/ were trapped in opened-cage \Cseventy/ with the yield 3:97 in the favour of species with one \Htwo/ per cage \cite{Murata2008JACS}.
The restoration of   closed cage retains approximately the same ratio of  \TwoHtwoatseventy/ to \Htwoatseventy/  \cite{Murata2008c70}.
 Molecules other than hydrogen trapped in opened-cage fullerenes are carbon monoxide  \cite{Iwamatsu2006}, water \cite{Iwamatsu2004,Xiao2007}, ammonia \cite{Whitener2008} and methane \cite{Whitener2009}.
Recently Kurotobi and Murata  succeeded in closing one of them and making the first  closed-cage endohedral complex with a trapped polar molecule, \Wateratsixty/ \cite{Kurotobi2011}.
The rotational modes of endohedral water were observed by inelastic neutron scattering (INS), far infrared (far-IR) and nuclear magnetic resonance (NMR) at cryogenic temperatures \cite{Beduz2012}.

Up to date \Htwoatsixty/ has been the most studied small molecule endofullerene.
The local environment of a hydrogen molecule in the fullerene cage  has a negligible inhomogeneous distribution of interaction parameters.
All the trapping sites are similar in  \Htwoatsixty/, except the isotopic distribution of carbon atoms and crystal field effects in  solid  \Htwoatsixty/. 
The \Htwoatsixty/  is a stable complex and can survive a short period of heating up to 500\,$^\circ$C under vacuum \cite{Murata2006}. 
These properties make \Htwoatsixty/ appealing for spectroscopic and theoretical investigations of  interactions between the molecular hydrogen  and carbon nanosurfaces.

Three spectroscopic techniques NMR, INS and infrared (IR),  have been used to study endohedral hydrogen. 
NMR studies cover spin lattice relaxation rates of \Htwoatsixty/ in organic solvents \cite{Sartori2006,Chen2010},
and in the presence of  paramagnetic relaxants \cite{Sartori_JACS2008,Frunzi2010}. 
NMR was used to follow the \ortho/-\para/ conversion in \Htwoatsixty/ in the presence of  molecular oxygen at 77\,K \cite{Turro2008} or upon photoexcitation of a \Cseventy/ triplet state \cite{Frunzi2011}.
NMR study of micro-crystalline \Htwoatsixty/ samples  at cryogenic temperatures shows splitting of the $J=1$ rotational state \cite{Carravetta2006,Carravetta2007}, a sign of the symmetry reduction from the icosahedral symmetry in the solid phase.
Similarly, splitting of the ground \ortho/ state was deduced from the  heat capacity measurements \cite{Kohama2009}.

The overview of  the low temperature NMR, INS and IR work on \Htwoatsixty/ is given by Mamone et al. \cite{Mamone2010CCR}.
The first IR study of  \Htwoatsixty/ was limited to  6\,K \cite{Mamone2009}.
The translational and rotational transitions appeared as sidebands to the hydrogen molecule  bond-stretching vibrational transition, $v=0\rightarrow 1$, in the mid-IR spectral range.
The direct translational and rotational transitions were not observed in the far-IR below 200\wn/ \cite{MinGe2011}.
The extension of IR studies to higher temperature made possible to probe the hydrogen-\Csixty/ interaction potential in the ground $v=0$ and first excited $v=1$ vibrational states and a whole range of hydrogen isotopologs \Htwo/, \Dtwo/ and HD was studied \cite{MinGe2011,Ge2011D2HD}.  
The isotope effects and translation-rotation coupling were also studied by INS in \Htwoatsixty/ and \HDatsixty/ \cite{Horsewill2010}.
The translational and rotational energies of \Htwoatsixty/ and \HDatsixty/   in the $v=0$ state determined by IR spectroscopy are consistent with  the low temperature INS results \cite{Horsewill2010}.
There is no Raman data on \Htwoatsixty/, except a  report on \Htwo/ inside an opened-cage fullerene \cite{Rafailov2005}.

In this paper  we will review the IR studies of hydrogen isotopologs in \Csixty/ and present the analysis of IR low temperature spectra of \Htwoatseventy/.
The far-IR properties of \Wateratsixty/ will not be reviewed here \cite{Beduz2012}.

%==========================================
\section{Theory\label{sec:theory}}
%==========================================
Quantum statistics plays an important role in the dihydrogen wavefunction symmetry and has a pronounced effect on the rotation of the hydrogen molecule \cite{Brown2003}.
The  symmetry relative to the interchange of two protons dictates that there are two forms of molecular hydrogen, called \para/- and \ortho/-\Htwo/.
The two proton spins ($I_p=1/2$) are in the anti-symmetric $I=0$ total nuclear spin state in \para/-\Htwo/ and in the symmetric $I=1$ state in \ortho/-\Htwo/.
Even rotational quantum numbers $J$  are allowed for \para/-\Htwo/ and odd $J$ for \ortho/-\Htwo/.
The nucleus of D is a boson, nuclear spin $I_d=1$.
Thus the rotational state with an even quantum number $J$ has \Dtwo/ in the state where the total nuclear spin of \Dtwo/ is either zero or two, $I=0,2$. 
This is called \ortho/-\Dtwo/, while \para/-\Dtwo/ has the total nuclear spin $I=1$ and odd $J$ values.

A homonuclear diatomic molecule with the total nuclear spin $I=1$ that is rotating in its ground state.
This   $J=1$ rotational state  is 118\wn/ for \Htwo/, and 58\wn/ for \Dtwo/, above the rotational ground state $J=0$ of even-$I$ species.
A thermal transition $J=1\rightarrow 0$ must be accompanied by the change of the total nuclear spin of the molecule, a process of very low probability.
The time constant of  thermal relaxation between the \ortho/ and the \para/ manifolds is very long and the room $T$ \ortho/-\para/ ratio is maintained even at cryogenic temperatures.
The equilibrium distribution of \Htwo/ nuclear spin isomers is $n_o/n_p=3$ 
and  of \Dtwo/ is $n_o/n_p=2$  at room temperature.
To change the total nuclear spin of a molecule  the two nuclei must experience  different magnetic field.
The \ortho/-\para/ conversion can be activated by using a paramagnetic center as a source of the magnetic field gradient. 
The equilibrium $n_o/n_p=1$ is reached  at 77\,K by  dispersing \Htwoatsixty/ on a zeolite surface and exposing it to  molecular oxygen which acts like a spin catalyst \cite{Turro2010}.
There are no \ortho/ and \para/ species for HD.
All rotational levels of HD are in thermal equilibrium and there  is one rotational ground state, $J=0$.

Quantum chemistry calculations are challenging for a hydrogen molecule in a weak van der Waals interaction with a large fullerene molecule.
The availability of experimental data on endohedral \Htwo/ has stimulated theoretical work in this direction.
Theoretical investigations currently cover the calculations of rotation-translation energies of hydrogen isotopologs in  \Csixty/  \cite{Xu2008,Xu2008HH_HD_DD} and  \Htwo/ in \Cseventy/ \cite{Xu2009}, and  the stability of \Csixty/ or \Cseventy/ with one or more  incarcerated \Htwo/ \cite{Lee2008,Sebastianelli2010}.
Empirical parameters of the Morse potential between H-H and contact Dirac interaction between H-C were adjusted \cite{Erkol2009} and 
density-fitting local M{\o}ller-Plesset theory  tested \cite{Dolgonos2011} using the experimental \Htwo/ vibrational frequency inside \Csixty/.
Classical molecular dynamics and density-functional theory have been combined to reproduce accurately the NMR chemical shift of $^1$H in \Htwoatsixty/ \cite{Oses2011}.

Both the hydrogen molecule and fullerene have closed shell electronic structures and therefore the interaction between them is the van der Waals interaction.
The simplicity makes \Htwoatsixty/  ideal for the studies of non-covalent bondings between \Htwo/ and carbon nano-surfaces, 
the  knowledge needed for  the    design of carbon-based hydrogen storage materials.
The high icosahedral symmetry of the \Csixty/ cavity is close to spherical and therefore \Htwoatsixty/ represents a  textbook example  of a body moving  in a  spherical potential well \cite{Shaffer1944,FLUGGE1971}.
In addition, \Htwo/   rotates around its center of mass.
\Htwo/ is not spherical and therefore its interaction with the walls of the cavity depends on its orientation what leads to the coupling between translational and rotational motion \cite{Cross2001}.
If  the translational and the rotational motions are coupled then in the spherical potential the conserved angular momentum is the sum of translational and rotational  angular momenta \cite{Yildirim2002,Xu2008}. 
\Htwoatsixty/ is a rare example where the quantum dynamics  of a diatomic rotor in a confined environment can be studied.
Another, but with limited degrees of freedom, interesting example related to the fullerenes is the quantum rotor C$_2$ in a metallofullerene C$_2$@Sc$_2$C$_{84}$ \cite{Krause2004,Michel2007}.
The two scandium atoms limit the translational motion and fix the rotational axis of  C$_2$ relative to the fullerene cage.
At low temperature the rotation of C$_2$ is hindered because it  has a small rotational constant and is therefore more susceptible to the corrugations of the carbon surface.
\Htwo/ provides examples of  two-dimensional  rotors, like \Htwo/ on a Cu surface \cite{Smith1996} or \Htwo/  in intercalated graphite \cite{Bengtsson2000}.

High pressure loading of solid \Csixty/ creates interstitial \Htwo/.
Exohedral \Htwo/ has been studied by IR \cite{FitzGerald2002,FitzGerald2006_PRB}, INS \cite{Kolesnikov1997,FitzGerald1999}, NMR \cite{Tomaselli2001,Tomaselli2003} and Raman \cite{Williams2002} spectroscopies.
Hydrogen is trapped in an interstitial site of octahedral symmetry and theory predicts translation-rotation coupling \cite{Yildirim2002,Herman2006}.
However, broadening of experimental lines has prohibited accurate determination of the \Htwo/-\Csixty/ interaction potential. 

The observed IR spectra of  hydrogen encapsulated in \Csixty/ consist of several absorption lines.
We construct a model Hamiltonian and a dipole moment operator with few adjustable parameters to describe accurately the position and intensity of such multi-line spectrum.

%==========================================
\subsection{Diatomic molecule in a spherical potential well\label{sec:diatomic_Hamiltonian}}
%==========================================

To describe the motion of a hydrogen molecule  inside \Csixty/ we use the following model.
The \Csixty/ is considered to be rigid, its center of mass is not moving  and it does not rotate.
We treat \Htwoatsixty/ as an isolated molecular complex and approximate the true icosahedral symmetry of an isolated    \Csixty/  with spherical symmetry.
It means that in this approximation \Htwo/ moves in a rigid spherically symmetric bounding potential provided by the cavity of \Csixty/.
Beside the translational movement inside \Csixty/  the hydrogen molecule has its internal degrees of freedom, vibration and rotation of two nuclei relative to its center of mass.
There are  no coupling terms  between \ortho/-  and \para/-states in our model Hamiltonian.

The theoretical work of Olthof et al. \cite{Olthof1996} is a comprehensive  description of the dynamics of a loosely bound molecule inside \Csixty/.
Olthof et al. model the intermolecular potential as a sum of atom-atom potentials and expand it in spherical harmonics.
They  determined the radial part of the wavefunction with discrete variable representation method.
The radial part of the wavefunction  in our approach is given by algebraic functions, solution of the three-dimensional spherical oscillator \cite{Shaffer1944,FLUGGE1971,Tannoudji1977}.
The advantage is that matrix elements are calculated in algebraic form avoiding time-consuming numerical integration.

\begin{figure}
\includegraphics[width=0.15\textwidth]{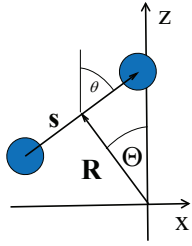}
\caption{\label{fig:coordinatesHHnoC} 
Coordinates  of a diatomic rotor  inside the \Csixty/ cage are the relative position $\svec/$ of two nuclei in the diatomic molecule  and    the diatomic center of mass displacement $\RCMvec/$ from  the \Csixty/ cage center.
In case of heteronuclear diatomic molecule (HD) the center of mass is not in the middle of the chemical bond between the two nuclei.
Two polar spherical coordinates, $ \thetaCM/ $ and $ \theta $, are shown.
(Online version in Colour)} 
\end{figure}

The position and orientation of the \Htwo/ molecule is given by spherical coordinates $\RCMvec/ =  \{\RCM/,\angularCM/ \}$, $\angularCM/=\{\thetaCM/,\phiCM/\}$  and  $\svec/ = \{s,\angularHtwo/ \} $, $\angularHtwo/=\{\theta,\phi \}$ where $\RCMvec/$ is the vector from the center of the \Csixty/ cage to the center of mass of \Htwo/ and $\svec/$ is the internuclear H-H vector, as shown in Fig.\,\ref{fig:coordinatesHHnoC}. 
%=============
The center of mass translational motion of \Htwo/ is given by eigenfunctions of the isotropic three-dimensional harmonic oscillator \cite{FLUGGE1971,Tannoudji1977}
\begin{equation}
\PsiTv_{NL\ML}(R,\angularCM/) = \psiTv_{NL}(R) \:\SpherHar{L\ML}(\angularCM/),
\end{equation}
where $\psiTv_{NL}$ is the  radial wave function  and $\SpherHar{L\ML}$ is the spherical harmonic. 
The size of the \Htwo/ molecule depends on its vibrational state $\ket{v}$.
Therefore both, the bounding potential and $\psiTv_{NL}(R)$, depend on the vibrational quantum number $v$.
The translational quantum numbers are  $N=0,1,2,\ldots$.
The orbital angular momentum quantum number takes values $L=N, N-2, \ldots,1$ or 0, depending on the parity of $N$, and the azimuthal quantum number is $\ML = -L, -L+1,\ldots,L$. 
The rotational wavefunctions, defined by the rotational quantum numbers $J=0,1,\ldots$ and $\MJ=-J,-J+1,\ldots, +J$, are given by the spherical harmonics $\SpherHar{J\MJ}(\theta,\phi)$.

We use bipolar spherical harmonics with overall spherical rank $\Lambda$ and component $\MLambda$
\begin{equation}\label{eq:ffunction_wave}
\Ffun{\Lambda\MLambda}{LJ}(\angularCM/, \angularHtwo/)=
	\sum_{\ML,\MJ}\Clebsch{L \ML J \MJ }{\Lambda \MLambda}\SpherHar{L\ML}(\angularCM/)\SpherHar{J\MJ}(\angularHtwo/),
\end{equation}
where  \Clebsch{}{} are  the Clebsch-Gordan coefficients  \cite{Tannoudji1977}. 
Then the  full wavefunction describing the motion of the \Htwo/ molecule  is
\begin{equation}\label{eq:wavefunction}
\ket{vJNL \Lambda \MLambda} = \psiV_v(s)\psiTv_{NL}(R)\Ffun{\Lambda\MLambda}{LJ}(\angularCM/,\angularHtwo/)
\end{equation}
where $\psiV_v(s)\equiv \ket{v}$ is the vibrational wavefunction with the quantum number $v$. 

The Hamiltonian $\H$ for the trapped molecule includes coupling terms between vibrational, translational, and rotational motion.  
For simplicity, we neglect all   matrix elements non-diagonal in  $v$ and introduce a parametric dependence  on $v$,  
\begin{eqnarray}\label{eq:Ham5d}
\H &=& \HVRv +\frac{p^2}{2 m} +\Vv(R,\angularCM/,\angularHtwo/),
\end{eqnarray}
where $\HVRv$ is the vibration-rotation Hamiltonian, $p$ is the molecular momentum operator and $m$ is the molecular mass of the diatomic molecule. 
$\Vv=\bra{v} V(\RCM/,s,\angularCM/,\angularHtwo/)\ket{v}$ is the potential energy of the hydrogen molecule in the vibrational state $\psiV_v(s)$. 
The vibration-rotation  Hamiltonian $\HVRv$ is diagonal in the basis set $\ket{vJNL \Lambda \MLambda}$ with eigenvalues given by 
\begin{equation}\label{eq:HamH2}
\EVRvJ =\hbar\wzeroV (v+ 1 / 2) +B_rJ(J+1),
\end{equation}
$B_r=B_e-\alpha_e(v+1/2) -D_eJ(J+1)$, where $\wzeroV$ is the fundamental vibrational frequency, $\alpha_e$ is the anharmonic correction and $D_e$ is the centrifugal correction to the rotational constant $B_e$ \cite{Herzberg1950,Dunham1932}.

We start from the general expansion of the potential in multipoles
\begin{equation}
\Vv(\RCM/,\angularCM/,\angularHtwo/) =\sum_{l,j,\lambda,\mlambda} \!\!\Vcoeff{\lambda m_\lambda}{lj}{v}(\RCM/) 
 \Ffun{\lambda \mlambda}{lj}(\angularCM/,\angularHtwo/).
\end{equation}
and expand the radial part of potential $ \Vcoeff{\lambda m_\lambda}{lj}{v}(\RCM/)$ in powers of $\RCM/$
\begin{equation}
\Vcoeff{\lambda m_\lambda}{lj}{v}(\RCM/) = \sum_n \Vcoeff{\lambda m_\lambda}{ljn}{v}\RCM/^n,
\end{equation}
where  $n \geq l$ and the  parity of $l$  and $n$ is the same.  

$V(\RCMvec/,\svec/)$ is a scalar and transforms under fully symmetric representation $A_g$ of the symmetry group $I_h$. 
The  spherical harmonics $\lambda=0,6, 10,\ldots$ transform like $A_g$ of the symmetry group $I_h$ \cite{Altmann1994}.
We use fully spherical approximation of the potential, $\lambda=0$.
Because $\lambda=0$  and   $\lambda = |l-j|, |l-j|+1,...,l+j$ it must be that  $l=j$.

The total potential is 
\begin{eqnarray}\label{eq:V_HD}
\Vv & = & \Vcoeff{00}{000}{v}\Ffun{00}{00} \nonumber \\ 
&& 
+  \left(  \Vcoeff{00}{002}{v} R^2 + \Vcoeff{00}{004}{v} R^4 \right) \Ffun{00}{00} 
+  \left(\Vcoeff{00}{111}{v}R +\Vcoeff{00}{113}{v} R^3\right) \Ffun{00}{11} 
+  \left(\Vcoeff{00}{222}{v}R^2  +\Vcoeff{00}{224}{v} R^4\right) \Ffun{00}{22},
\end{eqnarray} 
if we limit our expansion to $j=l=2$ and $n=4$.
The odd-$j$ terms are not allowed by symmetry for \Htwo/ and \Dtwo/  and thus  the coefficients \Vcoeff{00}{111}{v}  and \Vcoeff{00}{113}{v}  are zero for homonuclear diatomic molecules.

If we set  the constant off-set  $\Vcoeff{00}{000}{v}\Ffun{00}{00}$ to zero  and write the perturbation part as
\begin{equation}\label{eq:Vprime_HD}
\Vvprime =  \Vcoeff{00}{004}{v} R^4 \Ffun{00}{00}
 + \left(\Vcoeff{00}{111}{v}R +\Vcoeff{00}{113}{v} R^3\right) \Ffun{00}{11}
 +  \left(\Vcoeff{00}{222}{v}R^2  +\Vcoeff{00}{224}{v} R^4\right) \Ffun{00}{22},
\end{equation}
and  the isotropic harmonic term as
$\Vvzero =\Vcoeff{00}{002}{v} R^2 \Ffun{00}{00}$, then the total Hamiltonian reads
\begin{equation}\label{eq:HwithPerturbation}
\H = \HVRv +\frac{p^2}{2 m} +\Vvzero + \Vvprime.
\end{equation}

The unperturbed Hamiltonian eigenvalues in the basis $\ket{vJNL \Lambda \MLambda}$ are 
\begin{equation}
E^0_{vJNL\Lambda\MLambda }= \EVRvJ+\hbar\: \wzeroTv{v}(N+ 3 / 2),
\end{equation}
where $\wzeroTv{v}=\left(\Vcoeff{00}{002}{v}/(2\pi m)\right)^{1/2}$ is the frequency for translational oscillations within the cavity.

\begingroup
\squeezetable
\begin{table}[bt]
\begin{ruledtabular}
%...\SVN\NotebookC60\Notebooks\HamDiagTermAnalysis.nb
\caption{\label{tab:3D_Osc_Energies}
Translation-rotation energies of a 3D spherical oscillator  $\bra{vJNL\Lambda v}\H -\HVRv\ket{vJNL\Lambda}$ for a perturbation given by Eq.\,\ref{eq:Vprime_HD} and $\H$ by Eq.\,\ref{eq:HwithPerturbation} for few lower states.
Here 
${}^v\!\Delta^{004}= \frac{1}{^v\!\beta}\frac{15}{8}\frac{\Vcoeff{00}{004}{v}}{\Vcoeff{00}{002}{v}}$ ,
${}^v\!\Delta^{222}=\frac{\sqrt{5}}{20}\frac{\Vcoeff{00}{222}{v}}{\Vcoeff{00}{002}{v}} $,
${}^v\!\Delta^{224}= \frac{1}{^v\!\beta}\frac{7}{8\sqrt{5}}\frac{\Vcoeff{00}{224}{v}}{\Vcoeff{00}{002}{v}}  $,
and for  $^v\!\beta$ see Eq.\,\ref{eq:Rsquared}.
 }
\begin{tabular}{cl}
$JNL\Lambda$& $\bra{vJNL\Lambda v}\H -\HVRv\ket{vJNL\Lambda}$\\
\hline
&\\ % \hline
0000 &  $\hbar \wzeroTv{v}[\frac{3}{2}+{}^v\!\Delta^{004}]$\\
1001 &  $\hbar \wzeroTv{v}[\frac{3}{2}+{}^v\!\Delta^{004}]$\\
0111 &  $\hbar \wzeroTv{v}[\frac{5}{2}+\frac{7}{3}{}^v\!\Delta^{004}]$\\
&\\ % \hline
1110 &  $\hbar \wzeroTv{v}[\frac{5}{2}+\frac{7}{3}{}^v\!\Delta^{004}+10({}^v\!\Delta^{222}+{}^v\!\Delta^{224})]$\\
1111 &  $\hbar \wzeroTv{v}[\frac{5}{2}+\frac{7}{3}{}^v\!\Delta^{004}-5\ ({}^v\!\Delta^{222}+{}^v\!\Delta^{224})]$\\
1112 &  $\hbar \wzeroTv{v}[\frac{5}{2}+\frac{7}{3}{}^v\!\Delta^{004}+1\ ({}^v\!\Delta^{222}+{}^v\!\Delta^{224})]$\\
&\\ % \hline
2111 &  $\hbar \wzeroTv{v}[\frac{5}{2}+\frac{7}{3}{}^v\!\Delta^{004}+5\ ({}^v\!\Delta^{222}+{}^v\!\Delta^{224})]$\\
2112 &$\hbar \wzeroTv{v}[\frac{5}{2}+\frac{7}{3}{}^v\!\Delta^{004}-5\ ({}^v\!\Delta^{222}+{}^v\!\Delta^{224})]$\\
2113 & $\hbar \wzeroTv{v}[\frac{5}{2}+\frac{7}{3}{}^v\!\Delta^{004}+\frac{10}{7}\ ({}^v\!\Delta^{222}+{}^v\!\Delta^{224})]$\\
&\\ % \hline
0200 & $\hbar \wzeroTv{v}[\frac{7}{2}+5{}^v\!\Delta^{004}]$\\
0222 & $\hbar \wzeroTv{v}[\frac{7}{2}+\frac{21}{5}{}^v\!\Delta^{004}]$\\
&\\ % \hline
1201 & $\hbar \wzeroTv{v}[\frac{7}{2}+5{}^v\!\Delta^{004}]$\\
1221 &  $\hbar \wzeroTv{v}[\frac{7}{2}+\frac{21}{5}{}^v\!\Delta^{004}+7({}^v\!\Delta^{222}+\frac{9}{7}{}^v\!\Delta^{224})]$\\
1222 &  $\hbar \wzeroTv{v}[\frac{7}{2}+\frac{21}{5}{}^v\!\Delta^{004}-7({}^v\!\Delta^{222}+\frac{9}{7}{}^v\!\Delta^{224})]$\\
1223 & $\hbar \wzeroTv{v}[\frac{7}{2}+\frac{21}{5}{}^v\!\Delta^{004}+2({}^v\!\Delta^{222}+\frac{9}{7}{}^v\!\Delta^{224})]$\\

\end{tabular}
\end{ruledtabular}
\end{table}
\endgroup

The meaning of different parts of the perturbation is explained by their influence on  the energy levels of a harmonic 3D spherical oscillator, Table\,\ref{tab:3D_Osc_Energies}.
Translation-rotation coupling term $\Vcoeff{00}{222}{v}$ splits  energy of $\ket{vJNL\Lambda}$ state into  levels with different $\Lambda$, where $\Lambda= |L-J|,|L-J|+1, \ldots,L+J$.
For example,  the $N=L=J=1$ state is split into three levels with  different total angular momentum $\Lambda= 0,1,2$.
The ordering of levels depends on the sign of  $\Vcoeff{00}{222}{v}$.
The anharmonic correction to translation-rotation coupling  is $\Vcoeff{00}{224}{v}$. 
If  isotropic anharmonic  correction $\Vcoeff{00}{004}{v}$ is positive the distance between energy levels increases with $N$ and this correction is different for the levels with same $N$ but different $L$. 
For example, for positive $\Vcoeff{00}{004}{v}$ the $N=2, L=0$ level has higher  energy than $N=2, L=2$ level.

The length scale $^v\!\beta=m \wzeroTv{v}/\hbar$ (dimension m$^{-2}$) of the radial part of a 3D spherical oscillator wavefunction  is related to the expectation value of the center of mass amplitude in state $ \ket{N}$  as \cite{Shaffer1944} 
\begin{equation}\label{eq:Rsquared}
\bra{N} R^2\ket{N} = {}^v\!\beta^{-2}(N+3/2) = \hbar\sqrt{\frac{2\pi}{m \Vcoeff{00}{002}{v}}}(N+3/2).
\end{equation}
%C:\Users\Public\Documents\Samples\Fullerenes\Calc\AmplitudeOf3Doscillator.nb%

Terms described by the translation-rotation coupling coefficients  $\Vcoeff{00}{111}{v}$  and $\Vcoeff{00}{113}{v}$ do not appear in Table\,\ref{tab:3D_Osc_Energies} because the first order correction  to energies  vanishes as  the  matrix element  of $\Ffun{00}{11}$ is zero if diagonal in $L$ or $J$.
These terms   mix states with different $N$ and $J$.
For example, the  first excited rotational state $J=1$, $N=0$ (expectation value of HD center of mass is on the cage center) has the state $J=0$, $N=1$  (expectation value of HD center of mass is off the cage center) mixed in \cite{Ge2011D2HD}.
The effect is that HD is  forced to rotate about its geometric center instead of center of mass.

It was found by the 5D quantum mechanical calculation that the rotational quantum number $J$ is almost a good quantum number for the homonuclear \Htwoatsixty/ and \Dtwoatsixty/ and not for the heteronuclear \HDatsixty/ \cite{Xu2008HH_HD_DD}.
Indeed, $\Vcoeff{00}{22n}{v}$ mixes states with different $J$ for homonuclear species as well but the effect is reduced compared to the effect of $\Vcoeff{00}{11n}{v}$.
In the former case $J\pm 2$ and $L\pm 2$ are mixed while in the latter case the $J\pm 1$ and $L\pm 1$ states that have a smaller energy separation are mixed.

The states with different $\Lambda$ are not mixed in the spherical approximation, i.e. the total angular momentum $\mathbf{ \Lambda=L+J }$  is conserved and  $\Lambda$ is a good quantum number.
The other consequence of the spherical symmetry is that the  energy does not depend on  $\MLambda $.
Therefore it is practical to use a reduced basis and reduced matrix elements \cite{Gordy} which are independent of $\MLambda $. 
This reduces the number of states by  factor $2\Lambda+1$ for each $\Lambda$.

%==========================================
\subsection{Model Hamiltonian of \Htwoatseventy/ \label{sec:seventy_ham}}
%==========================================

A spherical approximation of the potential of a molecule trapped in   \Cseventy/ would be an oversimplification because of the elongated shape of \Cseventy/.
The symmetry of \Cseventy/ is $D_{5h}$, the distance between the centers of  two capping pentagons ($z$ direction) is 7.906\,\AA{ }.
%(7.968\,\AA \cite{Nikolaev1994})
The diameter of the equatorial   $xy$ plane is 7.180\,\AA \cite{Hedberg1997},
%(7.124\,\AA \cite{Nikolaev1994}).
similar to the  diameter of the icosahedral sphere of \Csixty/, 7.113\,\AA \cite{Hedberg1991}.
The anisotropy of the potential of \Htwo/ inside \Cseventy/  is supported by the 5D quantum mechanical calculation \cite{Xu2009}  what shows that the lowest translational excitation  in the $z$ direction is 54\wn/ and in the $xy$ plane is 132\wn/ while in \Csixty/ it is  180\wn/ and isotropic \cite{MinGe2011}.
We derive from the IR spectra (see below) that the $xy$ plane excitation energy is 151\wn/,  somewhat larger than  theoretically predicted.

Although the $z$ axis translational energy in \Cseventy/ is  three times less than  in the icosahedral \Csixty/, the effect of the \Cseventy/ potential on the rotational motion is moderate. 
The splitting of the $J=1$ state is 7\wn/ what is relatively small compared to the  rotational energy  120\wn/ in this  state \cite{Xu2009}.
%We expect the translation-rotation coupling of \Htwoatseventy/ be similar to \Htwoatsixty/, of the order of 10\wn/ \cite{Mamone2009,MinGe2011}.
 
To analyze the IR spectra of \Htwoatseventy/ we use a simplified Hamiltonian where the translational energy is represented in the form of the sum of two oscillators, 1D linear and 2D circular oscillator, and we do not consider anharmonic corrections   and the translation-rotation coupling.

The vibration-rotation energy  
\begin{equation}\label{eq:seventy_ham_vr}
\EVRvJz =\hbar\;\wzeroV (v+ \frac{1}{2}) +B_r^{(v)} J(J+1)+ \vkappa(3J_z^{2}-2),
\end{equation}
is the same as for \Htwoatsixty/ except the last 
 term which accounts for the axial symmetry of the \Cseventy/ potential with the rotational anisotropy parameter $\vkappa$ \cite{Yildirim2003}.
For example, the three-fold degenerate $J=1$ rotational state in $I_h$ symmetry  is split in $D_{5h}$ symmetry and if $\vkappa > 0$  the    $J_z=0$ state is $3\vkappa $ below the  twice degenerate $J_z=\pm 1$ rotational state.

The translational part is added to the vibration-rotation Hamiltonian, Eq.\,\ref{eq:seventy_ham_vr}, and the total energy reads
\begin{eqnarray}\label{eq:seventy_ham_twooscillators}
E^0_{vJJ_z;nln_z}&=& \EVRvJz + \hbar\; {}{^v\!\wzeroTxy}(n + 1)+\hbar\; {}{^v\!\wzeroTz}(n_z+ \frac{1}{2}).
\end{eqnarray}
Here the translational energy is written as a sum of two oscillators, a linear oscillator along the $z $ axis with translational quantum $ \hbar\: {}{^v\wzeroTz} $
and a 2D (circular)  oscillator \cite{FLUGGE1971,Tannoudji1977} in the $ xy $ plane with translational quantum $ \hbar\: {}{^v\wzeroTxy} $.
Quantum numbers $ n_z $ and $ n $ are positive integers including zero and 
$ l= n, n-2,\ldots,-n+2, -n $.

We will show below that the frequencies  of $z$ and $xy$ translational modes,   ${}{^1\!\wzeroTz} $ and ${}{^1\!\wzeroTxy} $, can be determined from the experimental data even though the translation-rotation coupling is not known.
We take the advantage of  translation-rotation coupling being  zero in the $ J=0 $ rotational state.
The complication rises from the fact that the potential is different  in the initial and final states of the IR transitions,   $v=0 $ and 1.
However, this complication could be resolved if the energy of the fundamental vibrational  transition $v=0\rightarrow 1$ (without change of $n$ and $n_z$) is known. 

The $ \Delta J=0 $ transition from the \para/-\Htwo/ ground state leads to two excitation energies in the IR spectrum, one for the $z$ mode and second for the  $xy$ mode
\begin{eqnarray}
E^0_{100;001}- E^0_{000;000} &=& \hbar[ \wzeroV+{}{^1\!\wzeroTz}+ \: ({}{^1\!\wzeroTxy}-{}{^0\!\wzeroTxy})+\frac{1}{2}\: ({}{^1\wzeroTz}-{}{^0\wzeroTz})]\label{eq:para_tr_energyZ},\\
E^0_{100;110}- E^0_{000;000} &=&  \hbar[ \wzeroV+{}{^1\!\wzeroTxy}+ \: ({}{^1\!\wzeroTxy}-{}{^0\!\wzeroTxy})+\frac{1}{2}\: ({}{^1\wzeroTz}-{}{^0\wzeroTz})]\label{eq:para_tr_energyXY}.
\end{eqnarray} 
Defining the fundamental \para/ transition energy as 
$
E^0_{100;000}- E^0_{000;000} =  \hbar 
[ 
\wzeroV 
+ 
({}{^1\!\wzeroTxy}-{}{^0\!\wzeroTxy})+\frac{1}{2}\: ({}{^1\wzeroTz}-{}{^0\wzeroTz}) 
]$
we may rewrite Eq.\,\ref{eq:para_tr_energyZ} and \ref{eq:para_tr_energyXY} as
\begin{eqnarray}
E^0_{100;001}- E^0_{000;000} &=& \hbar {}{^1\!\wzeroTz}+E^0_{100;000}- E^0_{000;000},\\
E^0_{100;110}- E^0_{000;000} &=&  \hbar {}{^1\!\wzeroTxy}+E^0_{100;000}- E^0_{000;000}\nonumber.
\end{eqnarray} 
From these equations translational frequencies in the excited $v=1$ state, ${}{^1\!\wzeroTz} $ and ${}{^1\!\wzeroTxy} $, can be determined without knowing the translation-rotation coupling.

The classification of energy levels up to $J=1$ and $n=n_z=1$  by irreducible representations  $\Gamma_i$ of the symmetry group  $D_{5h}$ is given in Table\,\ref{tab:Levels_organized_D5h}.
We get the irreducible representations $\Gamma_j$: $L=0\rightarrow A_1' $ and $L=1\rightarrow A_2'' + E_1' $ by subducting the translational states represented by spherical harmonics $\SpherHar{L\ML}$ from the full rotational group $O(3)$ to the symmetry group $D_{5h}$.
$A_1' $ is the \para/-\Htwo/ ground state, $n=l=n_z=0$.
The first excited state of the $z$ mode is $n_z=1 $ and $A_2''$.
The first excited state of the $xy$ mode $n=l=1$ is doubly degenerate $E_1' $.
The full symmetry when translations and rotations are taken into account  is $\Gamma_i=\Gamma_j\otimes\Gamma^{(J)}$.
For example, the \ortho/-\Htwo/ ground state, $J=1$ and $(nln_z)=(000)$, is split into 
$J_z=0$ ($A_2''$) and $J_z=\pm 1$ ($E_1'$), see Table\,\ref{tab:Levels_organized_D5h}.

\begingroup
\squeezetable
\begin{table}
\begin{ruledtabular}
\caption{\label{tab:Levels_organized_D5h}
Classification of energy levels  of \Htwo/ inside  the  cage of \Cseventy/ up to $J=1$ and $n=n_z=1$  by irreducible representations  $\Gamma_i$ of the symmetry group  $D_{5h}$. 
} 
\begin{tabular}{cc|l}
$J$ & $(nln_z)$ & $\Gamma_i$\\ 
\hline
0 & (000)& $A_1'$\\
1 & (000)& $A_2''$, $E_1'$\\
0 & (001)& $A_2''$\\
1 & (001)& $A_1'$, $E_1''$\\
0 & (110)& $E_1'$\\
1 & (110)& $A_1'$, $A_2'$, $E_2'$, $E_1''$
\end{tabular}
\end{ruledtabular}
\end{table}
\endgroup

%==========================================
\subsection{Induced dipole moment of hydrogen in spherical environment\label{sec:diatomic_dipole}}
%==========================================
IR light is not absorbed by vibrations and rotations of  homonuclear diatomic molecules \cite{Herzberg1950}. 
IR activity of \Htwo/ is induced  in the presence of intermolecular interactions, such as in the solid and liquid phases  \cite{Allin1955,Hare1955}, in constrained environments \cite{Hourahine2003,FitzGerald2002,FitzGerald2006_PRB,Herman2006}, and in pressurized gases \cite{Kudian1971,McKellar1974}.
IR spectra of such systems are usually broad due to inhomogeneities in the system or due to random molecular collisions.
As an exception, narrow lines are observed in semiconductor crystals \cite{Chen2002} and solid hydrogen \cite{Oka1993}.

An overview of collision-induced dipoles in gases and gas mixtures is given in the book by L. Frommhold \cite{Frommhold1993}.
The confinement of the endohedral \Htwo/ introduces two differences as compared to  \Htwo/ in the gas.
{\em First}, the translational energy of \Htwo/ is quantized.
In the gas phase it is a continuum  starting from zero energy.
{\em Second}, the variation of the distance between \Htwo/ and the carbon atom is limited  to the translational amplitude of \Htwo/ in the confining potential.
In the gas phase the distance varies from  infinity to the minimal distance given by the collision radius.
The selection rule $\Delta N = \pm 1$ for the endohedral \Htwo/ follows from these two conditions, as shown below.

Quantum mechanical calculations of induced dipoles are available for simple binary systems like \Htwo/-He, \Htwo/-Ar, and \Htwo/-\Htwo/.
An extensive set of theoretical results for the \Htwo/-He system associated with the roto-translation electric dipole transitions, both in the vibrational ground state $v=0$ and accompanying  the   $v=0\rightarrow 1$ transition of the \Htwo/ molecule, can be found in  \cite{Berns1978,Wormer1979,Meyer1986,Frommhold1987,Gustafsson2000}.
Related to the fullerene studies are  calculations of the dipole moment of CO@\Csixty/ \cite{Olthof1996} and exohedral \Htwo/ in solid \Csixty/ \cite{Herman2006}.

We express the induced part of the dipole moment as an interaction between hydrogen molecule and  \Csixty/.
Another approach was used  in \cite{MinGe2011} where the summation over 60 pair-wise induced dipole moments between \Htwo/ and  carbon atoms was done.
The relation between two sets of parameters was given \cite{Ge2011D2HD}.

We write the expansion of the dipole moment from the vibrational state $v$ to $v'$ in bipolar spherical harmonics and in power series of $\RCM/$ as
\begin{equation}\label{eq:expand_CM}
d_{v'v}^{}(\RCMvec/, \angularHtwo/)=\frac{4\pi}{\sqrt{3}}\sum_{l,j,n}\Acoeff{\lambda \mlambda}{ljn}{v'v}\RCM/^n \Ffun{\lambda \mlambda}{lj}(\angularCM/, \angularHtwo/).
\end{equation}
This is similar to the expansion of the potential discussed  above, except the dipole moment is a polar vector while the potential is a scalar.
The dipole moment transforms according to the irreducible representation $T_{1u}$ of the symmetry group $I_h$.
The spherical harmonics of the order $\lambda=1,5, 7,\ldots $ transform according to $T_{1u}$ of the symmetry group $I_h$ \cite{Altmann1994}.
We use  $ \lambda=1 $ and are interested in $v=0\rightarrow 1$ transitions.
In spherical symmetry it is sufficient to calculate one component of the dipole moment vector, $ \mlambda=0 $, and if we drop the explicit dependence of $d_{v'v}^{}$ on $v, v'$ and of \Acoeff{\lambda \mlambda}{ljn}{v'v} on $\lambda, \mlambda$, the $\mlambda=0$ component of the dipole moment reads
\begin{equation}\label{eq:expand_CM_Simple}
d_{0}^{}(\RCMvec/, \angularHtwo/)=\frac{4\pi}{\sqrt{3}}\sum_{l,j,n}\Acoeff{}{ljn}{}\RCM/^n \Ffun{10}{lj}(\angularCM/, \angularHtwo/).
\end{equation}

As   $\lambda = |l-j|, |l-j|+1,...,l+j$ and $\lambda=1$   it must be that  $l=|j\pm 1|$.
The possible combinations are $(lj) \in \{ (01), (10), (12), (21), (23),\ldots \}$. 
If we restrict the expansion up to $n=l=1$, we get
\begin{eqnarray}\label{eq:dipole_HD}
d_{0}^{}(\RCMvec/, \angularHtwo/)  & = & \frac{4\pi}{\sqrt{3}}\Acoeff{}{010}{}\Ffun{10}{01} \nonumber + 
\frac{4\pi}{\sqrt{3}} \left( \Acoeff{}{101}{}\Ffun{10}{10}+\Acoeff{}{121}{}\Ffun{10}{12} \right)\RCM/.
\end{eqnarray}

$\Acoeff{}{010}{}$ is zero for homonuclear diatomic molecules \Htwo/ and \Dtwo/. 
It is the sum of the HD permanent rotational dipole moment in the gas phase  and the induced rotational dipole moment inside \Csixty/.
The  selection rule is
$\Delta N = 0, \Delta J=0,\pm 1$, but $J = 0 \rightarrow 0$ is forbidden.

The expansion coefficients $\Acoeff{}{101}{}$ and $\Acoeff{}{121}{}$ are allowed by symmetry for both, homo- and hetero-nuclear diatomic molecules.
The expansion coefficient $\Acoeff{}{101}{}$ describes the induced dipole moment that is independent of the orientation of the diatomic molecular axis $\svec/$, selection rule 
$\Delta N = 0,\pm 1, \Delta J=0$, but $N = 0 \rightarrow 0$ is forbidden. 
$\Acoeff{}{121}{}$ describes the induced dipole moment that depends on the orientation of $\svec/$, 
$\Delta N = 0, \pm 1$, $\Delta J=0,\pm 2$, but $N = 0 \rightarrow 0$ and $J = 0 \rightarrow 0$ are both forbidden. 
All terms in Eq.\,\ref{eq:dipole_HD} satisfy the selection rule $\Delta \Lambda = 0, \pm 1$, but $\Lambda = 0 \rightarrow 0$ is forbidden. 

Although HD has a permanent rotational dipole moment, the induced dipole moment dominates inside \Csixty/ \cite{Ge2011D2HD}.

IR absorption line intensity\,\cite{MinGe2011} is proportional to the thermal population of the initial state.
If the  thermal relaxation between \para/-\Htwo/ and \ortho/-\Htwo/ is very slow we can define a temperature independent fractional \ortho/ and \para/ populations $n_k$ of the total number of molecules $\mathcal{N}$,  where $k=o, p$ selects \ortho/ or \para/-\Htwo/. 
Then the probability that the initial state $\ket{v_iJ_iN_iL_i\Lambda_i \MLambdai}$  is populated is 
\begin{equation}\label{eq:PopulationOP}
p_i = n_k  \frac{e^{-E_i/k_B T}}
{\sum_j g_je^{-E_j/k_B T} },
\end{equation}
where $E_i$ is the energy of the initial state measured from the ground state $v=N=0$  and $j$ runs over all \para/- (or  \ortho/-) \Htwo/ states in the basis used. 
$ g_j=2\Lambda_j+1 $ is the degeneracy of the energy level $E_j$.
Please note that $ g_i$ does not appear in the numerator because $p_i$ is the population of a individual state $\ket{v_iJ_iN_iL_i\Lambda_i \MLambdai}$ although the reduced basis  $\ket{vJNL\Lambda}$ is used.
Hetero-nuclear HD has no \para/ and \ortho/ species and the coefficient $n_k$ in Eq.\,\ref{eq:PopulationOP} must be set to one,   $n_k=1$.

%===========
% project file:  H2C70.opj
\begin{figure}
\includegraphics[width=0.6\textwidth]{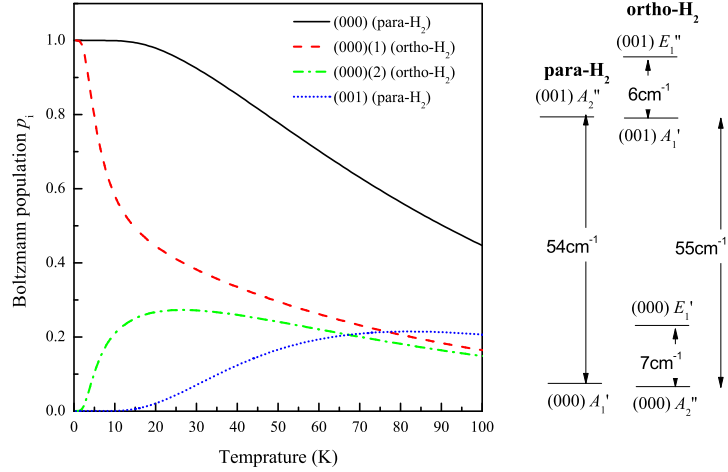}
\caption{\label{fig:OP_ground_Boltz} 
Boltzmann population of the \Htwoatseventy/ \para/ ground  translational state (000)(solid line) and first thermally excited  (dotted line)  state (001) and of the two  \ortho/ states, $A_2''$ (dashed line) and $E_1'$ (dash-dot line), in the ground translational state $(000)$, calculated using energy levels of the 5D quantum mechanical calculation \cite{Xu2009}.
Here  the \para/ states have $J=0$ and the \ortho/ states have $J=1$;
irreducible representations are from Table\,\ref{tab:Levels_organized_D5h}.
(Online version in Colour)} 
\end{figure}
%===========

%=========== 
% project file in the manscript folder:  H2C60spec.opj
%
\begin{figure}
\includegraphics[width=1\textwidth]{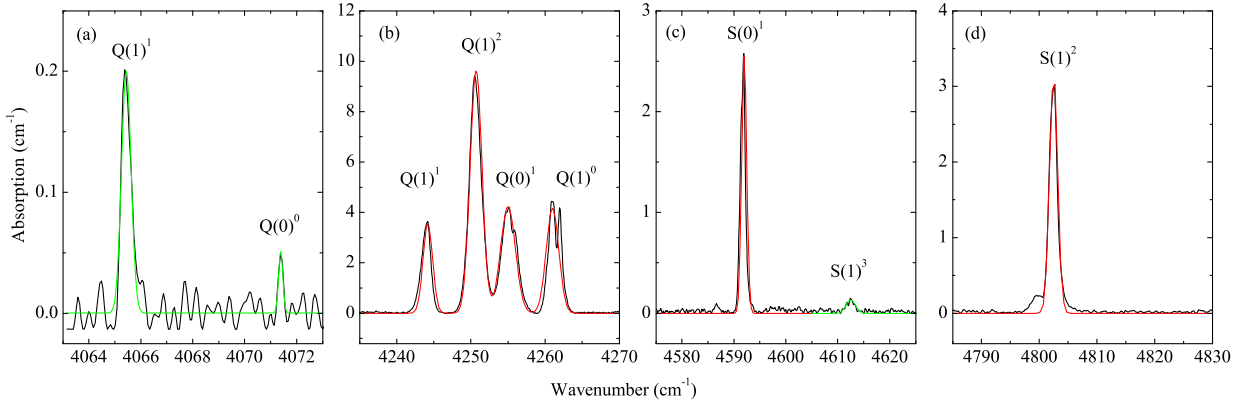}
\caption{\label{fig:Spec6KH2C60} 
Infrared absorption spectra of \Htwoatsixty/ at 6\,K are shown with the black line.
The green line in panel (a) is the Gaussian fit of fundamental \para/ and \ortho/ transitions, $\Delta N=\Delta J = 0$.
The red line in panels (b), (c), (d) is the simulated spectrum with parameters taken from the model fit of 200\,K spectra \cite{MinGe2011}. 
The $S(1)^3$ labeled green line in panel (c) is the Gaussian fit of the $\Delta N=0, \Delta J = 2$ forbidden \ortho/ transition, experimentally observed at 4612.5\wn/ and predicted to be at 4613.1\wn/ by our model \cite{MinGe2011}. 
Line labels are the same as in Fig.\,\ref{fig:OPLowT_transH2_HD_C60}a; the superscript to line label shows the final state $\Lambda$.
} 
\end{figure}
%===========

%==========================================
\subsection{Infrared absorption by \Htwoatseventy/ \label{sec:seventy_absorption}}
%==========================================

The infrared absorption in \Htwoatseventy/ is given by  electric dipole  operators $d_0$ and $d_{\pm 1}$, which transform like $A_2''$ and $E_1'$ irreducible representations of $D_{5h}$.
At low $T$ the symmetry-allowed transitions from the ground \para/ state  are
$   A_1' \stackrel{d_0}{\rightarrow} A_2''  $ and $   A_1' \stackrel{d_{\pm 1}}{\rightarrow} E_1'  $.
Two \ortho/ states, $A_2''$ and $E_1'$, are populated at low $T$, Fig.\,\ref{fig:OP_ground_Boltz}.
The transitions are $   A_2'' \stackrel{d_0}{\rightarrow} A_1'  $
and $   A_2'' \stackrel{d_{\pm 1}}{\rightarrow} E_1''  $.
The symmetry-allowed transitions from the $E_1'$ state are
$   E_1' \stackrel{d_0}{\rightarrow} E_1''  $ and
$   E_1' \stackrel{d_{\pm 1}}{\rightarrow} A_1', A_2', E_2'  $.

The IR absorption line area is proportional to the Boltzmann population of the initial state, Eq.\,\ref{eq:PopulationOP}.
The degeneracy  $g_j$ of the energy level $E_j$ is one or two in case of \Htwoatseventy/ .
Fig.\,\ref{fig:OP_ground_Boltz} shows the Boltzmann population
of the few first \para/ and \ortho/ levels in the $v=0$ vibrational state.
The \para/ state $(000)$ population (solid line) starts to drop above 20\,K as the first excited state $(001)$ (black dashed line) at 54\wn/ above the ground state becomes populated and drains the population from the $(000)$ state.
The abrupt change of the population of \ortho/ ground states above 2\,K is because of the transfer of population from the non-degenerate $A_2''$ ground level (dotted line) to the doubly degenerate $   E_1' $ (dash-dot line)  7\wn/ higher.
The  populations of $A_2''$ and $E_1' $ states decrease above 30\,K as the first excited translational $(001)$ \ortho/ state get more populated.

%===========
\begin{figure}
% project file:  C:\...\SVN\RSA_fullerene_2012\RSA_paper.opj
\includegraphics[width=0.4\textwidth]{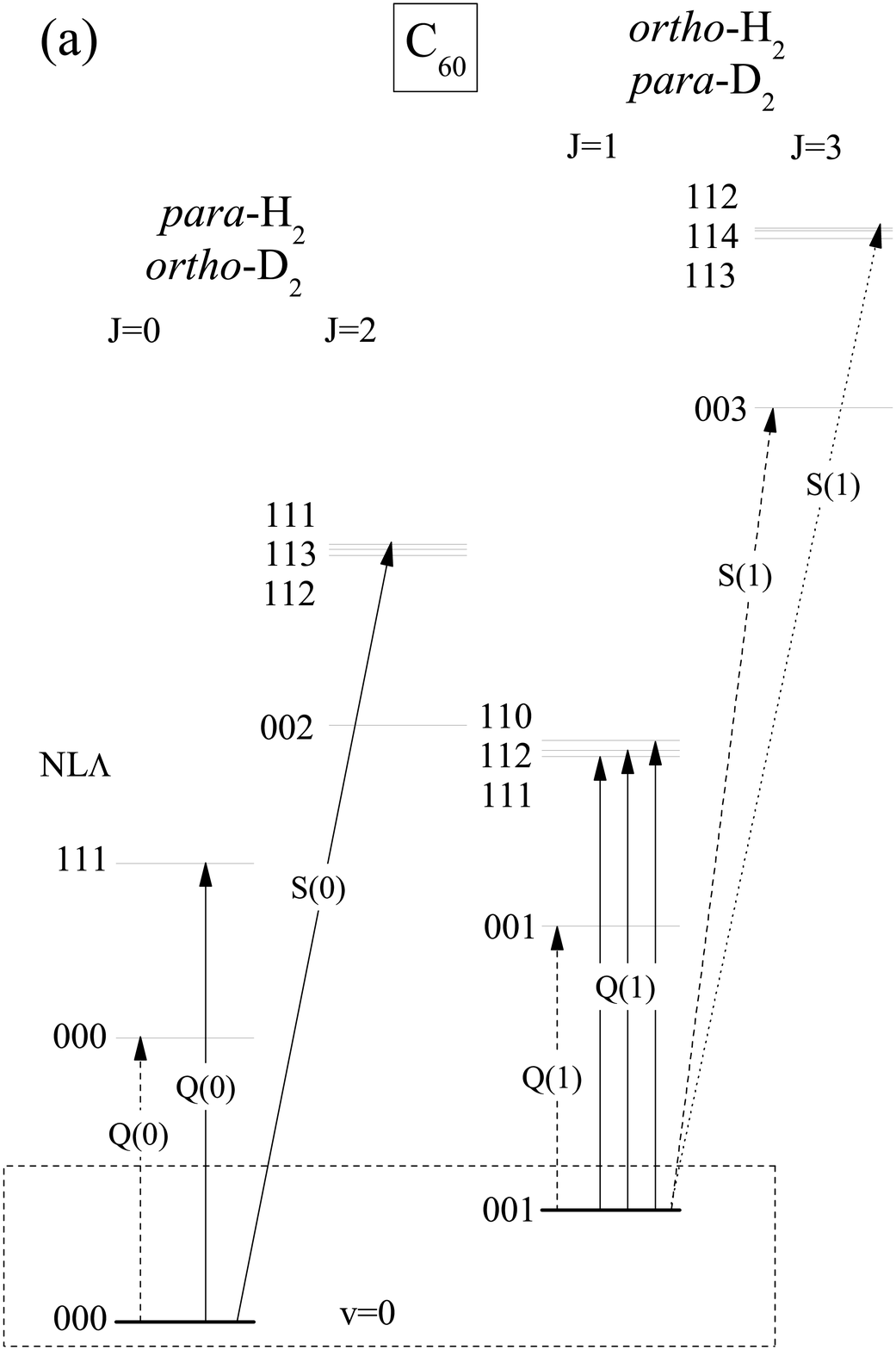}
\hspace{2cm}
% project file:  C:\...\SVN\RSA_fullerene_2012\RSA_paper.opj
\includegraphics[width=0.4\textwidth]{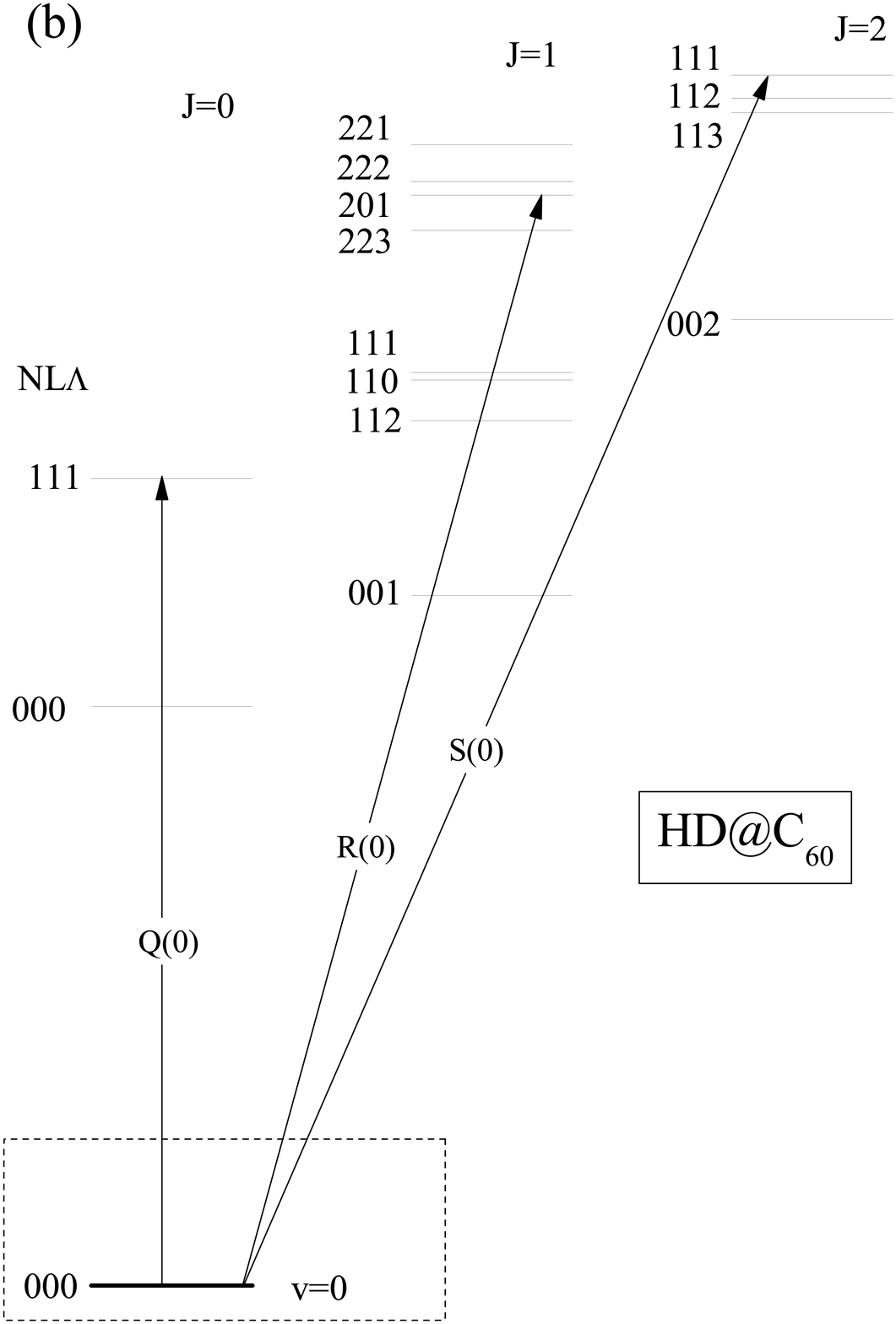}
\caption{\label{fig:OPLowT_transH2_HD_C60} 
Diagrams of the observed low temperature IR transitions in hydrogen isotopologs (a) \Htwo/, \Dtwo/ and (b) HD incarcerated in \Csixty/.
The initial states surrounded by a dashed box have the vibrational quantum number  $v = 0$; all the final states are in the excited vibrational state $v=1$. 
Dashed lines in (a) are forbidden transitions ($\Delta N=0$) that are observed in \Htwoatsixty/ but not in \Dtwoatsixty/.
The S(1) line (dotted) with $\Delta N =1$ is not observed in \Dtwoatsixty/.
The line label $Q(J_i)$ is used for $\Delta J=0$, $R(J_i)$ is used for $\Delta J=+1$ and $S(J_i)$ for $\Delta J=+2$ transitions where $J_i $ is the initial state $J$.
}
\end{figure}
%===========

%==========================================
\section{Experiment\label{sec:experiment}}
%==========================================

The endohedral complexes were prepared  by ``molecular surgery" as described in \, \cite{Komatsu2005,Murata2006}; 
\Htwoatsixty/, \Dtwoatsixty/  and \Htwoatseventy/ at Kyoto University,  and \HDatsixty/ at Kyoto University and Columbia University. 
The \HDatsixty/ sample was a mixture of the hydrogen isotopomers \Htwo/:HD:\Dtwo/ with  the ratio  45:45:10.
Since all \Csixty/ cages are filled, the filling factor for HD is $\rho=0.45$.
The content of \Cseventy/ sample was empty:\Htwo/:(\Htwo/)$_2$=28:70:2 and the  filling factor $\rho=0.7$. 
Experimental absorption spectra were corrected for the filling factor.

The \para/ enriched sample was made at Columbia University using molecular oxygen as a spin catalyst for  \ortho/-\para/ conversion \cite{Turro2008}.
Briefly, the \Htwoatsixty/ adsorbed on the external surface of NaY zeolite was immersed in liquid oxygen at 77\,K for 30 minutes, thereby converting the incarcerated \Htwo/ spin isomers to the equilibrium distribution at 77\,K,  $n_o/n_p=1$. 
The liquid oxygen was pumped away and  the endofullerene-NaY complex was brought back rapidly to room temperature. 
The \para/ enriched \Htwoatsixty/ was extracted from the zeolite with CS$_{2}$ and
% and the $^{1}$H-NMR spectrum of dissolved \Htwoatsixty/ was measured to confirm the \ortho/-\para/ conversion.% 
the solvent  was evaporated by argon.  
The  powder sample in argon atmosphere and  on dry ice arrived in Tallinn four days after the preparation. 

Powdered samples were pressed under vacuum into 3\,mm diameter pellets  for IR transmission measurements. 
Typical sample  thickness was 0.3\,mm. 
Two identical  vacuum tight chambers with Mylar windows were employed in the IR  measurements. 
The chambers were put inside an optical cold finger type cryostat with KBr windows.
In the measurements, the chamber containing the pellet for analysis was filled with He exchange gas while the empty chamber served as a reference.
Transmission spectra were obtained using a Bruker interferometer Vertex 80v with a halogen lamp and a HgCdTe or an InSb detector.
The apodized resolution was typically 0.3\wn/ or better. 

Transmission $ T_r(\omega)$ was measured as the light intensity transmitted by the sample divided by the light intensity transmitted by the reference.
The absorption coefficient $ \alpha(\omega) $ was calculated from the transmission $ T_r(\omega)$ through $\alpha(\omega)= -d^{-1} \ln\left[ T_r(\omega)(1-R)^{-2}\right]$,
with the reflection coefficient $ R=\left [ (\eta-1)/(\eta+1)\right ]^{2}$ calculated assuming a frequency-independent  index of refraction \cite{Homes1994},  $\eta=2$.
Absorption spectra were cut into shorter pieces around groups of \Htwo/ lines and a baseline correction was performed before fitting the \Htwo/ lines with Gaussians.

%==========================================
\section{Results and Discussion\label{sec:results}}
%==========================================

%==========================================
\subsection{\Csixty/}
%==========================================

The IR absorption spectra of \Htwoatsixty/ at 6\,K together with  the simulated spectra are shown in Fig.\,\ref{fig:Spec6KH2C60} and the diagram of energy levels involved in Fig.\,\ref{fig:OPLowT_transH2_HD_C60}a.
The simulated spectra are calculated  using Hamiltonian, dipole moment and the \ortho/-\para/ ratio parameters (see Table\,\ref{tab:parameters}) obtained from the fit  of  200\,K spectra \cite{MinGe2011}. 
Temperature does not affect these parameters and the line intensities follow the  Boltzmann population of initial states \cite{MinGe2011}.
Intensity of three lines, $Q(0)^0$ and $Q(1)^1$ in Fig.\,\ref{fig:Spec6KH2C60}\,a) and $S(1)^3$ in c)  can not be simulated because the  induced dipole moment theory does not describe $\Delta N=0$ transitions.
However, the position of these three lines was used to fit the Hamiltonian parameters \cite{MinGe2011}.
The splitting of the \ortho/ state $N=J=1$ into $\Lambda = 0, 1$ and 2 by translation-rotation coupling is seen in Fig.\,\ref{fig:Spec6KH2C60}\,b).
The \para/ $N=1, J=2$ and \ortho/ $N=1, J=3$ states are split into three sublevels as well.
However, because of the selection rule $\Delta \Lambda = \pm 1$ only one \para/ and one \ortho/ $S$-transition is IR active.

%==============
% project file: \Samples\Fullerenes\h2 at C60\OP_conversion\h2c60_OP_conversion.opj
%
\begin{figure}
\includegraphics[width=0.5\textwidth]{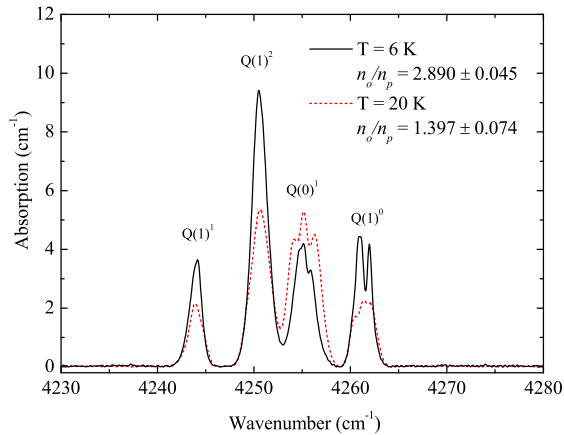}
\caption{\label{fig:Fig_OPconvLowT} 
Spectra in the region of $Q$ lines of a \para/ enriched \Htwoatsixty/ (dashed line)   and non-enriched sample  (solid line) at low $T$.
The thermal equilibrium \ortho/-\para/ ratio at 78\,K where the \para/ enrichment was done  is $n_o/n_p=1$.
(Online version in Colour)} 
% Sample arrived at Nov. 27. RT data (nov.28) was recorded right after midnight. Since we use average data, I assume it is about  12 hours after arrival, 4.5 days after preparation.
% 20K (Columbia) $n_o/n_p$ = 1.35 \pm 0.12
% 20K Murata $n_o/n_p$ = 2.8 \pm 0.2
% Nov.28 RT (Columbia) $n_o/n_p$ = 1.31 \pm 0.12
% Dec.16 RT (Columbia) $n_o/n_p$ = 2.74 \pm 0.27
% Murata RT $n_o/n_p$ =2.8 \pm 0.2
% For Murata sample, values taken from Low T paper.  
% 	ratios calclulated from fits in file Ratio.of OP_H2.opj
\end{figure}
%==============

The intensity of other lines in Fig.\,\ref{fig:Spec6KH2C60}\,b-d) is described accurately by two dipole moment parameters, $\Acoeff{}{101}{}$ and $\Acoeff{}{121}{}$, and \ortho/-\para/ ratio $n_o/n_p = 2.89\pm0.045$, what is very close to the statistical value three, Table\ref{tab:parameters}.
We confirmed the assignment of spectral lines to \para/- and \ortho/-\Htwo/ by measuring the spectrum of a \para/ enriched \Htwoatsixty/ sample \cite{MinGe2011}.
Low $T$ spectra in the region of $Q$ lines are shown in Fig.\,\ref{fig:Fig_OPconvLowT}.
The time delay between \para/ enrichment and first IR measurement was 4 days.
The 4255\wn/ \para/ line is stronger and other three \ortho/ lines are weaker in the \para/-enriched sample, as compared to the non-enriched sample. 

\begin{figure}
% project file:  \Samples\Fullerenes\D2 at C60\D2C60.opj
%
\includegraphics[width=0.6\textwidth]{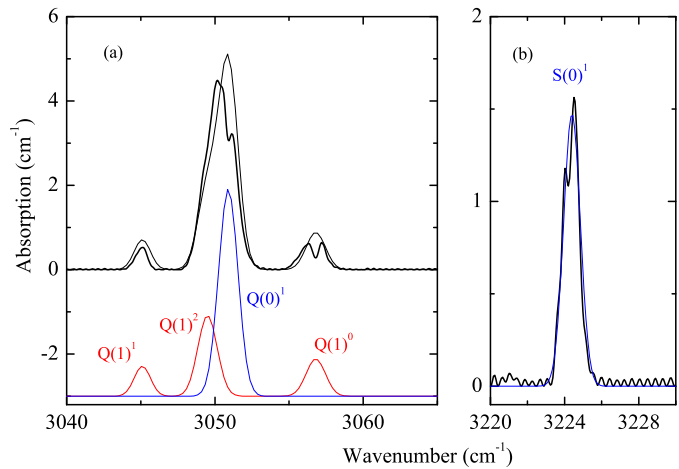}
\caption{\label{fig:FigSpec5KD2C60} 
Infrared absorption spectra of D$_{2}$@C$_{60}$ at 5\,K.
The simulated spectrum is shown by red line (\para/), blue line (\ortho/)
and black thin line (sum of \para/ and \ortho/ spectra).
The parameters for the simulated spectrum are taken from the model fit of the 90\,K spectrum \cite{Ge2011D2HD}.
Line labels are the same as in Fig.\,\ref{fig:OPLowT_transH2_HD_C60}a;  superscript to line label shows the final state $\Lambda$.
} 
\end{figure}

The Fig.\,\ref{fig:Fig_OPconvLowT} deserves some attention.
Lines $Q(0)^{1}$ and $Q(1)^{0}$ have a clear multicomponent structure  and it is different for the two samples.
This is not possible  in the approximation of spherical symmetry.
Even not by considering the true icosahedral symmetry of \Csixty/ because the lowest $\Lambda $ value for the state which is split by the icosahedral symmetry is three while the initial and final states of the optical transitions under discussion have only $\Lambda=0,1 $.
Later measurements on the relaxed but initially \para/ converted sample showed that the difference between the line shapes of the normal and   \para/ converted sample is not due to the different \ortho/-\para/ ratio.
Thus it is likely that the difference in the line splitting is due to a different impurity content of the two samples. 
However, it is not completely excluded that the crystal field or the distortion of the \Csixty/ cage  is responsible for the part of this splitting.
Note that the $Q(1)^{0}$ line of \Dtwoatsixty/,  Fig.\,\ref{fig:FigSpec5KD2C60}, bears similar splitting pattern as in \Htwoatsixty/.

The \Dtwoatsixty/ spectrum,  Fig.\,\ref{fig:FigSpec5KD2C60}, is shifted to lower frequency compared to \Htwoatsixty/ because of the heavier mass of \Dtwo/.
The spectrum has less lines than the \Htwoatsixty/ spectrum.
The missing transitions in  \Dtwoatsixty/ are shown  in red color  in Fig.\,\ref{fig:OPLowT_transH2_HD_C60}a and they  belong to $J$-odd  species   which are minority for \Dtwo/.

The splitting of $N=J=1$ into $\Lambda = 0, 1$ and 2 states is similar in \Dtwo/ and \Htwo/.
The magnitude of the splitting, $\hbar \wzeroTv{v} \frac{\sqrt{5}}{20}\frac{\Vcoeff{00}{222}{v}}{\Vcoeff{00}{002}{v}}$, see Table\,\ref{tab:3D_Osc_Energies},
is less in \Dtwo/ because $\wzeroTv{1}$  is smaller although  $\Vcoeff{00}{002}{1}$ and  $\Vcoeff{00}{222}{1}$ are similar for  two isotopologs, Table\,\ref{tab:parameters}.
Line  $Q(1)^2$  overlaps partially with $Q(0)^1$ line, Fig.\,\ref{fig:FigSpec5KD2C60}.
However, this is not because the translation-rotation splitting  is different for two isotopologs but because of a smaller anharmonic correction $\alpha_e$ of the rotational constant for \Dtwo/, Table\,\ref{tab:parameters}.

%========================================================
%===========

\begin{figure}
% project file:  ..\Samples\Fullerenes\HD at C60\Graphs_for_paper_corrected_for_filling_factor\HD_pictures_for_the_HD_D2_paper.opj
%
\includegraphics[width=0.6\textwidth]{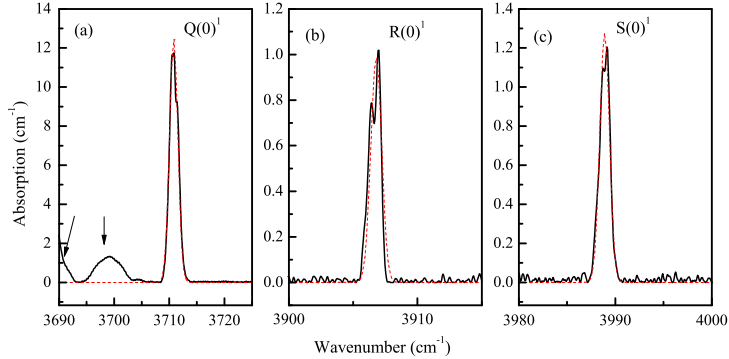}
\caption{\label{fig:FigSpec5KHDC60} 
Infrared absorption spectra of HD@C$_{60}$ at 5\,K, black line.
The parameters for the simulated spectrum (red line) are taken from the model fit of 90\,K spectra \cite{Ge2011D2HD}.
The broad lines indicated by arrows are not due to HD.
Line labels are the same as in Fig.\,\ref{fig:OPLowT_transH2_HD_C60}b;  superscript to line label shows the final state $\Lambda$.
(Online version in Colour)} 
\end{figure}

The spectrum of \HDatsixty/ is more simple, Fig.\,\ref{fig:FigSpec5KHDC60}, because  there are no \para/ and \ortho/ species and therefore only  one  state is populated at low $T$.
There is one spectral line, not present in homonuclear dihydrogen, labeled $R(0)$ in Fig.\,\ref{fig:OPLowT_transH2_HD_C60}b and Fig.\,\ref{fig:FigSpec5KHDC60}.
HD has no inversion symmetry and the ban on $\Delta J =1 $ transitions is lifted.
The classification of this transition as $\Delta J =1 $ is arbitrary first, because the weight of $ JNL\Lambda=1201 $ is only 0.5 in this  state \cite{Ge2011D2HD} and secondly, the change of translational state is $\Delta N= 2$, not allowed by any of the dipole operator components in our model.
%What state is mixed in?
%Eigenstate = 252,  Eigenvalue = 5870.8
 %ODDj states,  QuantumNumbers = {v,N,L,J,\[CapitalLambda],M\[CapitalLambda]}
  %  Pure state = 277,  amplitude = -0.08174,    {1,1,1,0,1,0}
   % Pure state = 278,  amplitude = -0.35552,    {1,3,1,0,1,0}
    %Pure state = 279,  amplitude = -0.09698,    {1,5,1,0,1,0}
 %   Pure state = 280,  amplitude = -0.00894,    {1,0,0,1,1,0}
 %   Pure state = 282,  amplitude = -0.7075,    {1,2,0,1,1,0}
 %   Pure state = 283,  amplitude = -0.1802,    {1,2,2,1,1,0}
 %   Pure state = 285,  amplitude = -0.1339,    {1,4,0,1,1,0}
 %   Pure state = 286,  amplitude = 0.02724,    {1,4,2,1,1,0}
 %   Pure state = 288,  amplitude = 0.0045,    {1,6,0,1,1,0}
 %   Pure state = 289,  amplitude = 0.01762,    {1,6,2,1,1,0}
 %   Pure state = 290,  amplitude = -0.5429,    {1,1,1,2,1,0}
 %   Pure state = 292,  amplitude = 0.04678,    {1,3,1,2,1,0}
 %   Pure state = 293,  amplitude = 0.0375,    {1,3,3,2,1,0}
 %   Pure state = 295,  amplitude = 0.03044,    {1,5,1,2,1,0}
 %   Pure state = 296,  amplitude = 0.00458,    {1,5,3,2,1,0}
 %   Pure state = 298,  amplitude = 0.07808,    {1,2,2,3,1,0}
 %   Pure state = 300,  amplitude = 0.00462,    {1,4,2,3,1,0}
 %   Pure state = 303,  amplitude = -0.00188,    {1,6,2,3,1,0}
 %   Pure state = 305,  amplitude = 0.00182,    {1,3,3,4,1,0}
 %Dominant pure state = 282,    {{1,2,0,1,1,0},0.5006}
The next component in weight 0.29 in the final state is $ JNL\Lambda=2111 $ which makes this transition $\Acoeff{}{121}{}$-active,  $N=0\rightarrow 1, J=0\rightarrow 2$ and therefore it would be more appropriate to classify it as an $S$ line. 
%C:\Users\Public\Documents\Samples\Fullerenes\Calc\Ham5DElementsReduced.nb
%Line positions have very little $T$ depndence below room temperature \cite{MinGe2011,Ge2011D2HD}.This tells us that the dihydrogen-\Csixty/ interaction potential is $T$ independent, likely because    the  spherical shape and the radius of \Csixty/ do not depend on $T$.
Another interesting feature of the HD spectrum is the absence of $ \Delta N=0, \Delta J= 1  $ ($ 0000\rightarrow 1001 $) transition although the induced dipole moment coefficient $\Acoeff{}{010}{}$ gives a two orders of magnitude larger dipole moment than the permanent dipole moment of free HD, as was discussed in \cite{Ge2011D2HD}.
This $\Delta J=1$ transition in \HDatsixty/ is suppressed because there is an interference of two dipole terms, $\Acoeff{}{010}{}$  and $\Acoeff{}{101}{}$ which have opposite signs.
The final state consists of  80\% of the pure rotational state $J=1, N=0$ and 18\% of the the pure translational state $N=1, J=0$.
This mixed final state has matrix elements from  the ground state for both $A$ coefficients, $\Acoeff{}{010}{}$  and $\Acoeff{}{101}{}$, which nearly cancel each other.

The observed low $T$ IR transitions of \Htwo/, \Dtwo/ and \HDatsixty/ are collected in Table\,\ref{tab:LowT_transitionsH2D2HD}.
The content of the unperturbed $\ket{JNL\Lambda}$ state in the final state is above $|\xi^1|^2=0.9$ in homonuclear species in most cases while for HD it varies and could by as low as 0.5.
The mixing of states  $\xi^v$ is proportional to the energy separation of mixed states, $E_j-E_i$ in the first order perturbation theory.
$\Ffun{00}{00}$    couples states where $L_i=L_j$ and $J_i=J_j$. 
These states are far from each other as $L_i=L_j$ only if $N_j=N_i\pm 2$. 
The other term $\Ffun{00}{22}$ mixes $L_j=L_i\pm2$ and $J_j=J_i\pm2 $ what are even further from each other for small $N$. 
It was found by Xu et al. \cite{Xu2009} that $J$ is almost a good quantum number in \Htwo/ and \Dtwoatsixty/ but not in \HDatsixty/.
Indeed, the distance between states mixed decreases  for HD as  $\Ffun{00}{11}$ mixes $L_j=L_i\pm 1$ and $J_j=J_i\pm 1 $ and for this $N_j=N_i\pm 1$.

%=========== 

\begingroup
\squeezetable
\begin{table*}
\begin{ruledtabular}
\caption{\label{tab:LowT_transitionsH2D2HD} 
Experimental  IR absorption line parameters  of endohedral dihydrogen isotopologs in \Csixty/ at  5\,K. 
$\omega$ is the transition frequency in wavenumber units, \wn/ and $S$ is the integrated absorption line area in \area/ units.
$J N L \Lambda$ are the quantum numbers  of the main component with the weight $|\xi^1|^2$ in the final state $v=1$.
The $|\xi^0|^2$ of the main component $J N L \Lambda$ in the initial $v=0$ state for  \Htwo/  are 0.97 of 0000 and 0.97 of 1001, for \Dtwo/ are 0.99 of 0000 and 0.99 of 1001, and for  HD is 0.95 of 0000.
Integrated absorption cross-section $\sigma$ is used by some authors. The integrated line are $S$ is transformed into  $\sigma$ (unit cm/molecule) using
$\sigma=S V/ \mathcal{N}$
where $\mathcal{N}$ is the number of hydrogen molecules in the volume $V$ \cite{Loudon1983}. 
Here  $\mathcal{N}/V=1.48\times 10^{21}$\,cm$ ^{-3} $, the number density of molecules in solid \Csixty/.
* It was not possible to separate overlapping $Q(0)^1$ and $Q(1)^2$ lines of \Dtwo/ and the line area given is the sum of both transitions.
}
\begin{tabular}{ccccccccccc}
 &  \multicolumn{3}{c}{\Htwo/ }&\multicolumn{3}{c}{\Dtwo/ } &\multicolumn{3}{c}{HD }&Final state, $v=1$\\
\cline{2-4}\cline{5-7}  \cline{8-10}\cline{11-11}
&&&&&&&&&&\\
Line label & $\omega$ 	& $S$&  $|\xi^1|^2$  	& $\omega$ 	& $S$&   $|\xi^1|^2$ 	& $\omega$ 	& $S$&$|\xi^1|^2$  & $J N L \Lambda$ \\
\hline
&&&&&&&&&&\\

$Q(1)^1$&	4065.4&	 0.093 &  0.98	& --	 & --&	0.99 &-- & -- &0.80 &1001	\\
$Q(0)^0$	&4071.4	&0.011		&0.98	&--&--&0.99&--&--&0.96&0000	\\
$Q(1)^1$	&4244.5	&5.76&0.95	&3045.1	&	0.98&0.97&--&--&0.94&1111	\\
$Q(1)^2$	&4250.7	&19.3&	0.94&3050.2*	&	10.8&0.96&--&--&0.64&1112	\\
$Q(0)^1$	&4255.2	&10.6&	0.94&3050.2*	&	10.8&0.96& 3710.8& 23.5&0.73&0111	\\
$Q(1)^0$	&4261.0	&8.86&0.94	&3056.8	&	1.55&0.96&--&--&0.64&1110	\\
$R(0)^1$	&--	&--	&0.61	&--	&--&0.61& 3906.7& 1.05&0.50&1201	\\
$S(0)^1$	&4592.0	&3.08	&0.94&3224.35 & 1.66	&0.96&3988.9&1.6&0.54&2111\\
$S(1)^1$	&4612.5	&0.3	&0.97&--	&--	&0.96&--&--&0.58&3003\\
$S(1)^2$	&4802.5	&5.65	&0.94&--	&--	&0.86&--&--&0.64&3112	\\
\end{tabular}
\end{ruledtabular}
\end{table*}
\endgroup

The vibrational frequency $\wzeroV$ is redshifted  from its gas phase value for \Htwo/ and \Dtwo/, Table\,\ref{tab:parameters}.
The relative change of the frequency 
$ [\wzeroV (\mathrm{gas})- \wzeroV (C_{60})] / \wzeroV (\mathrm{gas})  $ 
depends on the cage and is independent of the hydrogen isotopomer  \cite{Buckingham1960}.
Based on our fit results, Table\,\ref{tab:parameters}, we get that for the \Htwo/ $ \wzeroV (C_{60})/\wzeroV (\mathrm{gas}) = 0.9763$ and for the HD $ \wzeroV (C_{60})/\wzeroV (\mathrm{gas}) = 0.9773$. 
We used the average of these two ratios to calculate the \Dtwoatsixty/ fundamental vibrational frequency, $ \wzeroV=2924 $\,\wn/ what was necessary to fit the IR spectra to a model Hamiltonian \cite{Ge2011D2HD}.
At this point we cannot say how much of the redshift is caused by change in the zero-point vibrational energy and how much is caused by  the change of  anharmonic corrections  to the  vibrational levels in  the \Csixty/ as our data set is limited to energy differences of $v=0$ and $v=1$ levels only.
Note that the vibrational frequency of \Htwoatsixty/ is $\wzeroV = 4062.4$\wn/, Table\,\ref{tab:parameters}, while the \para/ vibrational transition is shifted up  by  9\wn/, to $ \omega_0=4071.4 $\wn/, Table\,\ref{tab:LowT_transitionsH2D2HD}.
This is because the zero-point translational energies are different in the $ v=0 $ and $ v=1 $ vibrational states \cite{MinGe2011}.

\begingroup
\squeezetable
\begin{table*}
\begin{ruledtabular}

\caption{\label{tab:parameters}
Values of the fitted parameters  for \Htwoatsixty/ \cite{MinGe2011} and for HD and \Dtwoatsixty/ \cite{Ge2011D2HD}.
The vibrational and rotational constants of a free molecule of the three hydrogen isotopologs are shown for comparison; gas phase $ \wzeroV $ is calculated including all terms $(v+1/2)^{k}$ up to $k=3$ \cite{Huber1979}.
 The parameter $\Vcoeff{00}{224}{v}$ is set to zero for  \HDatsixty/ and \Dtwoatsixty/.
 Parameter errors are given in\,\cite{Ge2011D2HD}.
} 
\begin{tabular}{cccccccc}
   \multicolumn{1}{l}{$\kappa_i$} 
 & \multicolumn{2}{c}{\Htwoatsixty/}  						& \multicolumn{2}{c}{HD@C$_{60}$}  		& \multicolumn{2}{c}{D$_2$@C$_{60}$} & \multicolumn{1}{r}{Unit}  
 \\  
\hline    
& $v=0$                     & $v=1$                       
& $v=0$             		& $v=1$             
& $v=0$               		& $v=1$              & \\
%\cline{2-7}
 
$\Vcoeff{00}{002}{v}$ 
& $14.28$				& $15.95$ 	       
& $16.36$ 				& $17.48$  
& $16.78$ 				& $16.46$    &Jm$^{-2}$
\\
$\Vcoeff{00}{004}{v}$ 
& $2.21 \cdot 10^{21}$		& $2.192 \cdot 10^{21}$ 
& $1.88 \cdot 10^{21}$		& $2.13 \cdot 10^{21}$ 
& $2.16 \cdot 10^{21}$ 		& $2.39 \cdot 10^{21}$ &Jm$^{-4}$
\\
$\Vcoeff{00}{222}{v}$        
& $0.563 $			& $1.20 $	       
& $1.31 $ 			& $2.04 $
& $1.1 $			& $1.4 $      &Jm$^{-2}$
\\
$\Vcoeff{00}{224}{v}$ 
& $2.21	\cdot 10^{20}$		& $1.03 \cdot 10^{20}$
& 0  & 0 
& 0  & 0    &Jm$^{-4}$
\\
$\Vcoeff{00}{111}{v}$
&--- 	&---                                               
& $3.11	\cdot 10^{-10}$ 	& $3.25	\cdot 10^{-10}$
&---	&---	&Jm$^{-1}$
\\
$\Vcoeff{00}{113}{v}$
&--- 	&---
& $4.3 \cdot 10^{10}$    	& $7.1	\cdot 10^{10}$
&---	&---	&Jm$^{-3}$ 
\\\\
$ n_o/n_p $ 
& \multicolumn{2}{c}{	$2.89 $} 
& \multicolumn{2}{c}{---} 
& \multicolumn{2}{c}{	$2$} 	& 
\\
\Acoeff{}{010}{}
& \multicolumn{2}{c}{---} 
& \multicolumn{2}{c}{$-1.5	\cdot 10^{-32}$}
& \multicolumn{2}{c}{---} & Cm
\\
\Acoeff{}{101}{}
& \multicolumn{2}{c}{$9.1 \cdot 10^{-22}$} 
& \multicolumn{2}{c}{$7.0 \cdot 10^{-22}$} 
& \multicolumn{2}{c}{$6.9 \cdot 10^{-22}$} & C
\\
\Acoeff{}{121}{}
& \multicolumn{2}{c}{$-4.3	\cdot 10^{-22}$}
& \multicolumn{2}{c}{$-2.7	\cdot 10^{-22}$}
& \multicolumn{2}{c}{$-2.9	\cdot 10^{-22}$} & C
\\
\\
%\cline{2-7} 
& 	\Htwoatsixty/       & H$_2$
&	HD@C$_{60}$    		& HD
&	D$_2$@C$_{60}$      & D$_2$ & \\
\cline{2-7}

$\wzeroV$
& $ 4062.4 $ 		& 4161.18
& $ 3549.7 $  	& 3632.20       
& $	2924 $  				& 2993.69 & \wn/ 
\\
$B_e$
& $ 59.87 $		& 60.853
& $ 45.4 $		& 45.655
& $ 29.89 $		& 30.444 & \wn/
\\
$\alpha_e$
& $ 2.97 $		& 3.062
& $ 1.70 $		& 1.986
& $ 1.09 $		& 1.079  & \wn/
\\
$D_e$
& $ 4.83 \cdot 10^{-2}$		& $ 4.71  \cdot 10^{-2}$
& $ 1.5 \cdot 10^{-1}$		& $ 2.605 \cdot 10^{-2}$
& $ 8 \cdot 10^{-3}$		& $ 1.141 \cdot 10^{-2}$ & \wn/ \\

\end{tabular}
\end{ruledtabular}
\end{table*}
\endgroup

The rotational constant $B_e$,   the vibrational correction $\alpha_e$, and the centrifugal correction $D_e$ of the hydrogen  inside \Csixty/ and in free space are compared in  Table\,\ref{tab:parameters}. 
The smaller than the gas phase value of $B_e$ may be interpreted as 0.8\% (0.9\%) stretching of the nucleus-nucleus distance $s$ in \Htwoatsixty/ (\Dtwoatsixty/), as $B_e\sim \langle 1/\!s^{2}\rangle$.
An attractive interaction between hydrogen atoms and the cage causes $s$ to be longer.
The elongation of the equilibrium proton-proton distance  is consistent with the redshift of $\wzeroV$ \cite{VandeWalle1998}.
However, the anharmonic vibrational correction to rotational constant, $\alpha_e$,  is smaller inside the cage than in the gas phase.
Here the cage has the opposite, repulsive, effect and reduces the elongation of the proton-proton distance in the excited $v$ states when compared to \Htwo/ being in the free space.
This is supported by the fact that within the error bars  $\alpha_e$ of \Dtwoatsixty/ is equal to $\alpha_e$ of \Dtwo/.
The vibrational amplitude of \Dtwo/ is less than of the \Htwo/ and therefore the repulsive effect of the cage becomes important at $v>1$ for \Dtwo/.

Among the rotational and vibrational constants of \HDatsixty/ the centrifugal correction $D_e$  to the rotational constant is anomalously different from its gas phase value compared to other two isotopomers, Table\,\ref{tab:parameters}.
Positive $D_e$ means that the faster the molecule rotates, the longer is the bond.
We speculate that since the rotation center of HD inside the cage is further away from the deuteron, the centrifugal force on the deuteron increases and the bond is stretched more than in the free HD molecule.

A similar system to the one studied here is  exohedral \Htwo/ in \Csixty/.
There \Htwo/ occupies  the octahedral interstitial site in  the \Csixty/ crystal. 
The prominent features in the exohedral \Htwo/ IR spectra \cite{FitzGerald2006_PRB} are  the translational, $\Delta N=\pm 1$, sidebands to the fundamental transitions, $\Delta v=1$ and $\Delta J=0, 2$.
The redshift of the fundamental vibrational frequency is about 60\wn/, which is less than in \Htwoatsixty/ where it is 98.8\wn/. 
Also the separation of translational $N=0$ and $N=1$ levels, approximately 120\wn/,  is smaller, when compared to 184.4\wn/ in \Htwoatsixty/.
It is likely that the main contribution to the latter difference comes from the larger van der Waals volume available for \Htwo/ in the octahedral site than in the \Csixty/ cage.

%==============
% project file:  H2C70_filling_factor_corrected_EXP_spectra.opj
\begin{figure}
\includegraphics[width=1\textwidth]{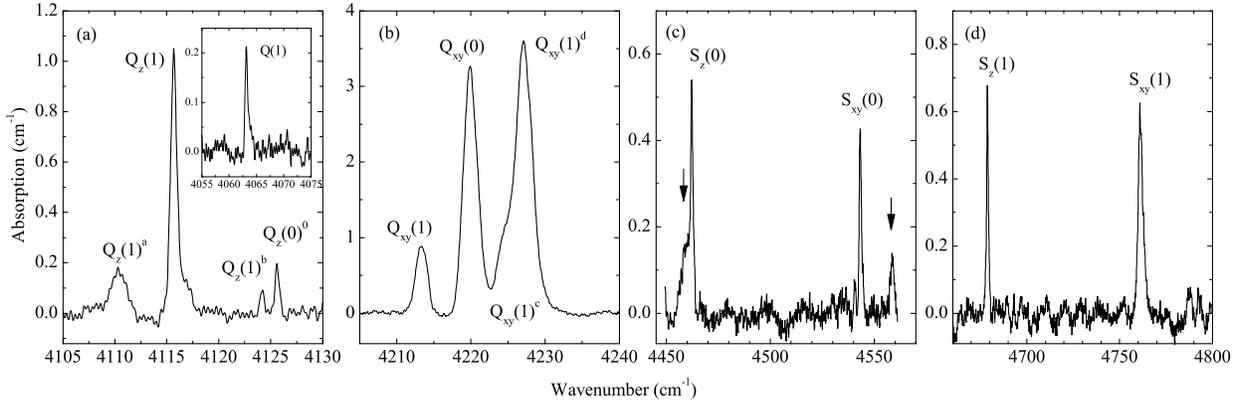}
\caption{\label{fig:H2C70spectra5K} 
IR absorption spectra of \Htwoatseventy/ at 5\,K. Features shown with arrows in panel (c) are likely caused by impurities, not by endohedral \Htwo/.
} 
\end{figure}
%===========

\begin{figure}
\includegraphics[width=0.4\textwidth]{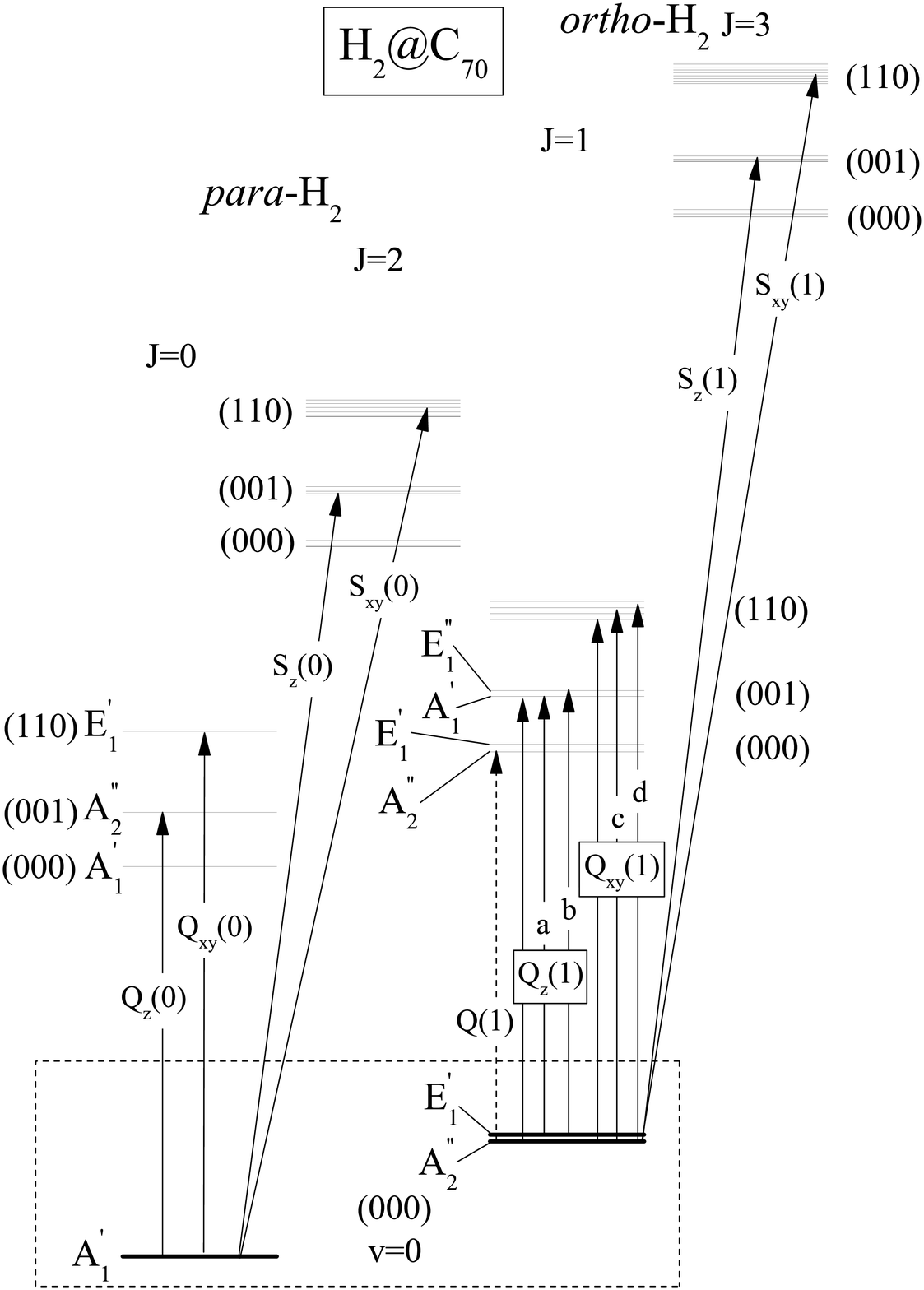}
\caption{\label{fig:OPLowT_transC70} 
Energy levels and observed IR transitions of \Htwoatseventy/   from the ground  \para/ and \ortho/ translational states in the $v=0 $ state to the excited vibrational state $v=1 $.
The irreducible representations of the symmetry group $D_{5h}$ are given for few lower states and are listed in Table\ref{tab:Levels_organized_D5h}.
The ordering of the states is based on \cite{Xu2009}.
\ortho/-\Htwo/ transitions marked by a and b are $Q_z(1)^a$ and $Q_z(1)^b$ in Table\,\ref{tab:OPLowT_transitionsC70} and start from the thermally excited state, which is $E_1'$ $(000)$ according to \cite{Xu2009}.
There are three experimentally observed \ortho/-\Htwo/ transitions from the ground $A_2''$ $(000)$ state:  $Q_{xy}(1)$, $Q_{xy}(1)^c$, and $Q_{xy}(1)^d$.
} 
\end{figure}
%==============

%==========================================
\subsection{\Htwoatseventy/}
%==========================================

IR absorption spectra of \Htwoatseventy/ measured at 5\,K are shown in Fig.\,\ref{fig:H2C70spectra5K} and the scheme of energy levels with the low $T$ transitions indicated with arrows is shown in Fig.\,\ref{fig:OPLowT_transC70}.
The transitions where $\Delta J=0$ and the translational state changes by $\Delta n = 1$ or $ \Delta n_z = 1 $ are labeled as $Q_{xy}(J)$ and  $Q_z(J)$ where $J$ is the rotational quantum number of the initial state. 
The transitions where $\Delta J=2$ are labeled by $S_{xy}(J)$ and  $S_z(J)$.
The $z$ and $xy$ translational modes have distinctly different energies.
For the $Q$ lines they are shown in  different panels, Fig.\,\ref{fig:H2C70spectra5K}\,a  and \ref{fig:H2C70spectra5K}\,b.
The \para/ and \ortho/ $S$ lines are well separated because of different rotational energies.
The \para/ $S$ lines are shown in Fig.\,\ref{fig:H2C70spectra5K}\,c and \ortho/ $S$ lines are shown in Fig.\,\ref{fig:H2C70spectra5K}\,d.
The $Q$ lines cannot be sorted out into \para/ and \ortho/ that easily. 
An exception is the $Q(1)$ transition, inset to Fig.\,\ref{fig:H2C70spectra5K}\,a. 
This is a pure vibrational transition, $v=0\rightarrow 1$, without the change of translational or rotational states.
The corresponding (fundamental) \para/ transition $Q(0)$ would be at  4069.3\wn/ according to the model, Table\,\ref{tab:OPLowT_transitionsC70}. 
We expect it to be  much weaker than the \ortho/ $Q(1)$ transition, as  in \Htwoatsixty/, and for this reason it is not observed in the experiment.

%===========
% project file:  H2C70.opj
\begin{figure}
\includegraphics[width=0.32\textwidth]{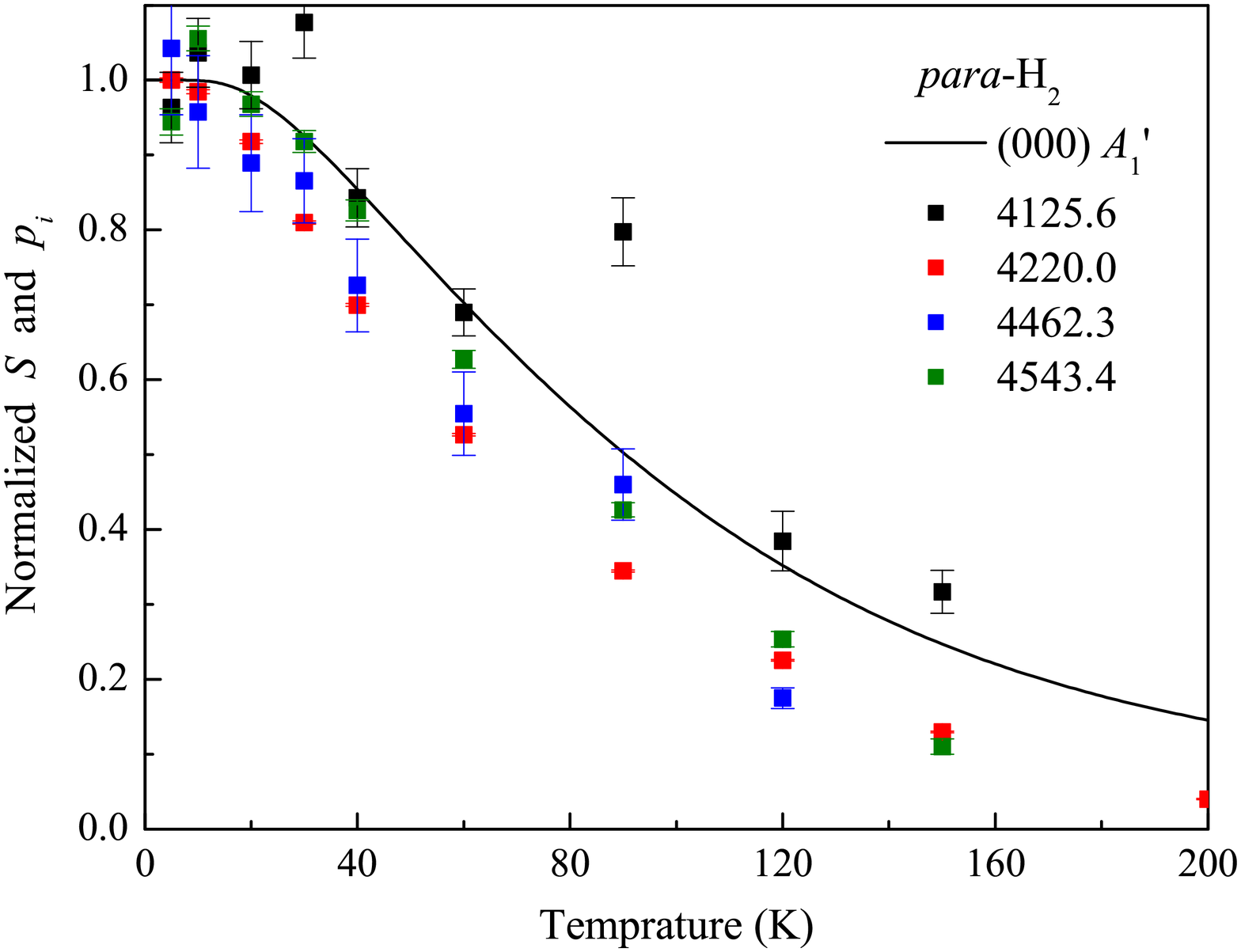}
\includegraphics[width=0.32\textwidth]{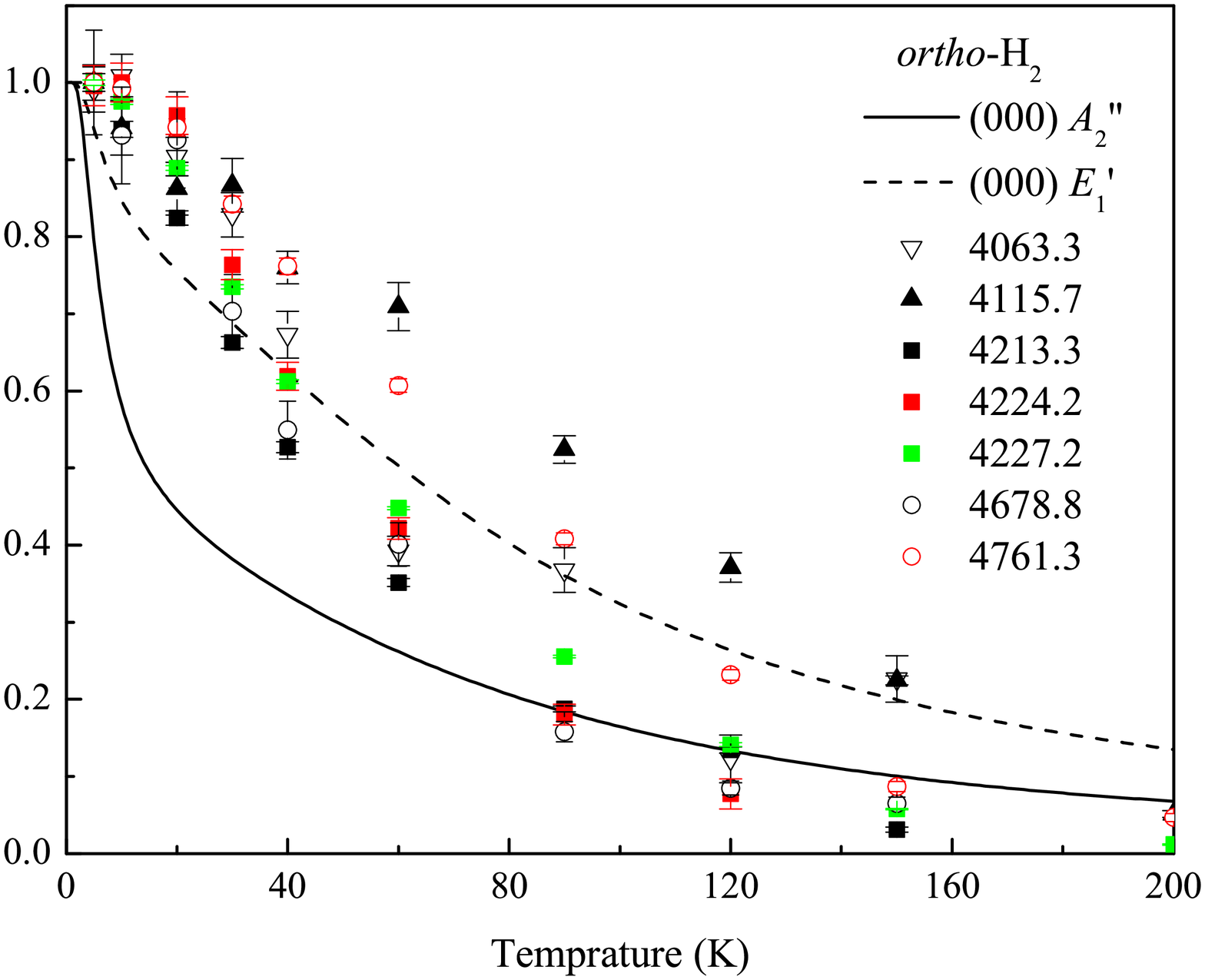}
\includegraphics[width=0.32\textwidth]{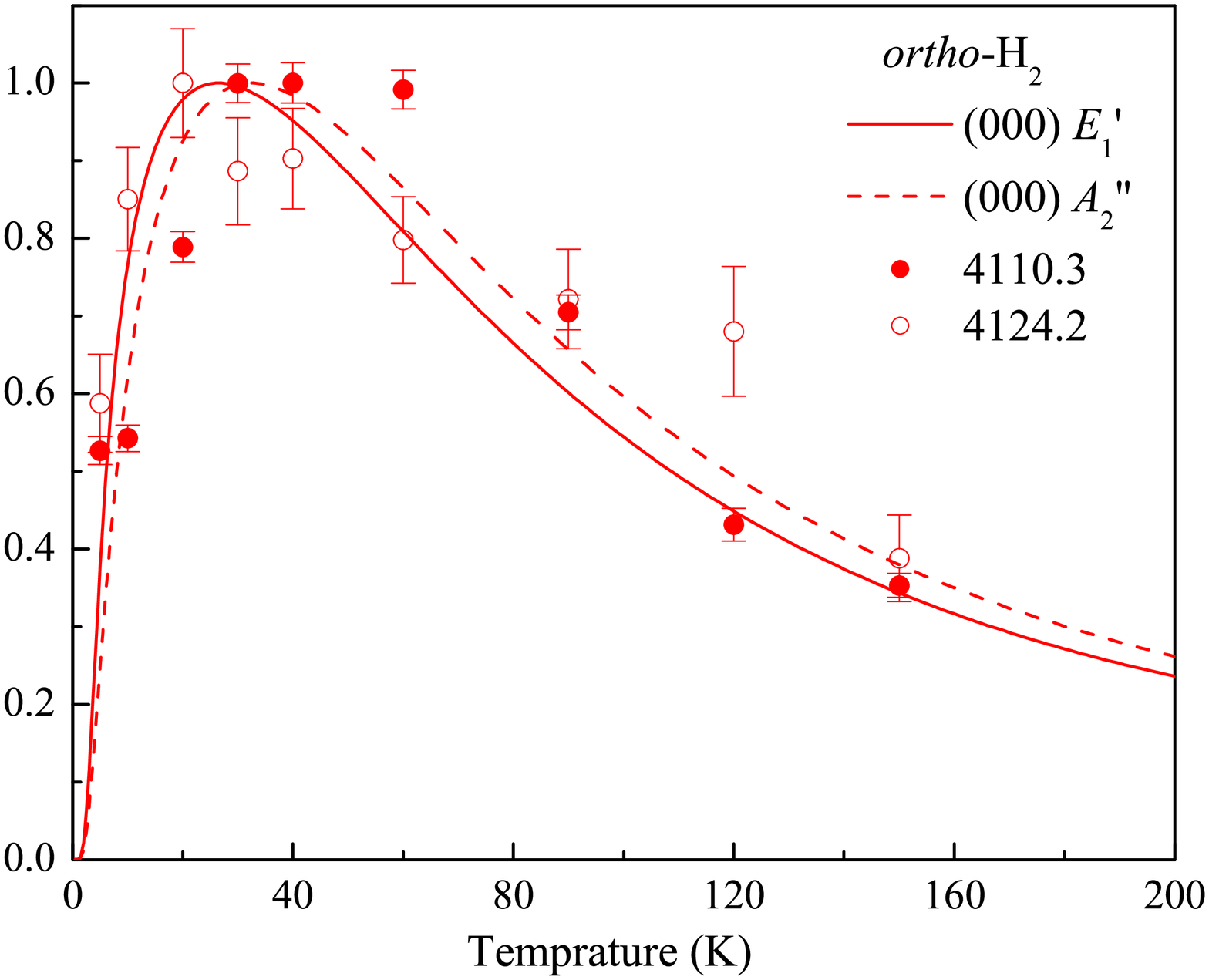}
\caption{\label{fig:C70TdepIntensity} 
Normalized line area and Boltzmann population of \para/ and \ortho/ transitions of \Htwoatseventy/ starting from the ground translational state.
(a) \para/ lines, (b) and (c)    \ortho/ lines. 
Solid line in (b) is the normalized intensity when the initial state is the ground state  $A_2''$ ($J_z=0$) and in (c) when the initial state is $E_1'$ ($J_z=\pm1$), 7\wn/ above $A_2''$.
This corresponds to  $\vkappa>0$, Eq.\,\ref{eq:seventy_ham_vr}. 
Dashed line in (b) and (c) is the  reversed situation when the ground state is doubly degenerate $E_1'$ and   $A_2''$ is 7\wn/ above it, i.e.  $\vkappa<0$.
The normalized intensities are calculated using the translational energy levels from the 5D quantum mechanical calculation \cite{Xu2009} and Eq.\,\ref{eq:PopulationOP}.
}
\end{figure}
%===========

In Fig.\,\ref{fig:C70TdepIntensity} we plot the normalized line area and normalized Boltzmann population.
Fig.\,\ref{fig:C70TdepIntensity}\,a shows the $T$ dependence of \para/ lines.
The assignment of the 4125.6\wn/ line to $Q_z(0)$ and the 4219.9\wn/ line to $Q_{xy}(0)$ is supported by our model that is summarized in Table\,\ref{tab:OPLowT_transitionsC70} and will be discussed below.

The uniaxial symmetry of \Cseventy/ splits the \ortho/ ground state $J=1$. This splitting is about 7\wn/$\approx 10$\,K and creates a sharp $T$ dependence of  the \ortho/ line intensity at low $T$ as is shown in Fig.\,\ref{fig:OP_ground_Boltz}.
This feature could be used to determine which of the $Q$ lines belong to the \ortho/ species.
Two lines, 4110.5\wn/ and 4124.2\wn/, have a $T$ dependence consistent with  the \ortho/ transitions starting from the ground translational $(000)$ state of $   E_1' $ symmetry, Fig.\,\ref{fig:C70TdepIntensity}\,a.
However, according to our model  there is no transition close to 4124.2\wn/ from the thermally excited \ortho/ $   E_1' $ state, Table\,\ref{tab:OPLowT_transitionsC70}.
This might be due to our simplified model where the rotational anisotropy parameter ${}^{v}\kappa$ is assumed equal in $v=0$ and  $v=1$ vibrational states.
The rest of the \ortho/ lines are shown in Fig.\,\ref{fig:C70TdepIntensity}\,b.
Several lines follow the same $T$ dependence, the fundamental transition at 4063.2\wn/, $Q_{xy}(1)$ lines 4213,4, 4224.7, and 4226.9\wn/ and the $S_{z}(1)$ line at 4678.8\wn/.
Lines $Q_{z}(1)$ at 4116 and $S_{xy}(1)$ at 4761\wn/ have distinctly different $T$ dependence than the rest of the lines.
Such deviation could be due to the overlap of transitions starting from $A_2''$ {\em and} $E_1'$ symmetry states.
However, none of the ortho lines behaves like the theoretical prediction, black line, so the possibility  that the ground ortho state is doubly degenerate $E_1'$ and not the $A_2''$, i.e. ${}^{0}\kappa<0$, is not ruled out by our data. 
The dashed black line shows the ground state population if the ground state is $E_1'$ and  the $A_2''$ state is 7\wn/ above it.
This $T$ dependence is more close to the experimental situation than the solid black line, where ${}^{0}\kappa>0$.
The change of sign of ${}^{0}\kappa$ affects intensities only a little if the transitions start from the thermally excited state, Fig.\,\ref{fig:C70TdepIntensity}\,c, where the red dashed line is the normalized Boltzmann population of $A_2'' $ state {\em if} this state is 7\wn/ above the ground state $   E_1' $, i.e. if ${}^{0}\kappa<0$.

It is not unreasonable that ${}^{0}\kappa<0$ and the $ E_1' $ ($J_z=\pm 1$) state is the ground rotational state  of \ortho/-\Htwo/.
It requires that there is an attraction instead of repulsion between H and C when \Htwo/ is in the center of the \Cseventy/ cage.
The attraction is stronger for the \Htwo/ if its molecular axis is in the $xy$ plane, $J_z=\pm 1$,  where the distance between H an C is less than if the axis is along the $z$, $J_z=0$, where the H-C distance is longer. 
Indeed, the attraction between C and H  in \Htwoatsixty/ was deduced from the redshift of vibrational energy and reduced rotational constant compared to \Htwo/   in the gas phase \cite{MinGe2011}.
However, it was found by 5D quantum mechanical calculation that $   A_2'' $ is the ground state \cite{Xu2009}.
More elaborate model with dipole moment parameters that describe the IR line intensities together with the Boltzmann population could resolve the issue. 

\begingroup
\squeezetable
\begin{table*}
\begin{ruledtabular}
% C70modelHam.nb

\caption{\label{tab:OPLowT_transitionsC70} 
Low temperature IR transitions of \Htwoatseventy/ observed at 5\,K from the ground vibrational state $v=0$ to the excited vibrational state  $v=1$.
$J (nln_z)$ are the quantum numbers for the initial and final states and the labels indicate corresponding transitions in Fig.\,\ref{fig:H2C70spectra5K} and \ref{fig:OPLowT_transC70}.
The experimental spectra were fitted with Gaussians, $S_{exp}$ is the experimental line area;  the experimental line position $\omega_{exp}$ could always be determined with a precision better than 0.1 \wn/.
The model  parameters, Eq.\ref{eq:seventy_ham_twooscillators}, were  determined as described in the text by matching the model frequency  $\omega_{mod}$ to the experimental frequency $\omega_{exp}$ for four lines indicated with error {\bf 0}. 
The  model parameters are vibrational frequency $\wzeroV=4069.1$\wn/, 1D oscillator frequency along the $z$ axis  $^v\!\wzeroTz=56.5$\wn/, 2D circular oscillator frequency $^v\!\wzeroTxy=150.9$\wn/,  and the rotational anisotropy parameter $\vkappa=3.1$\wn/.
The rotational constant  $B_e=59.865$\wn/ and its corrections, $\alpha_e=2.974$\wn/ and  $D_e=0.04832$\wn/ were  assumed to be the same as in \Htwoatsixty/.
}

\begin{tabular}{ccccccclc}
\multicolumn{2}{c}{Frequency } & $S_{exp}$ & \multicolumn{2}{c}{Initial }&\multicolumn{2}{c}{Final } &Label & Error\\
\multicolumn{2}{c}{(\wn/)} & (\area/) & \multicolumn{2}{c}{$v=0$ }&\multicolumn{2}{c}{$v=1$ } &  & (\wn/)\\
\cline{1-2}\cline{4-5}  \cline{6-7}\cline{9-9}
&&&&&&&&\\ 
$\omega_{exp}$ & $\omega_{mod}$&   & $J$  & $(nln_z)$ & $J$  & $(nln_z)$ && $\omega_{exp}-\omega_{mod}$\\
\hline
&&&&&&&&\\ 
4063.2& 4063.2	&$0.146 \pm 0.005$ 	& 1& (000) & 1 & (000) & $Q(1)$		&\bf 0\\
--& 4069.1&-- 	& 0					& (000) & 1 & (000) & $Q(0)$		&--\\
%\hline
&&&&&&&&\\
4110.3& 4110.3	&$0.311 \pm 0.014$	& 1& (000) & 1 & (001) & $Q_z(1)^a$	&\bf 0\\
4115.7& 4119.7	&$0.824 \pm 0.009$	& 1& (000) & 1 & (001) & $Q_z(1)$	&-4.0\\
4124.2& 4119.7 	&$0.045 \pm 0.007$	& 1& (000) & 1 & (001) & $Q_z(1)^b$	&4.5\\
4125.6& 4125.6	&$0.096 \pm 0.007$	& 0& (000) & 0 & (001) & $Q_z(0)$	&\bf 0\\
%\hline
&&&&&&&&\\
4213.3& 4214.1	&$1.759 \pm 0.024$	& 1& (000) & 1 & (110) & $Q_{xy}(1)$&-0.8\\
4220.0& 4220.0	&$7.702 \pm 0.027$	& 0& (000) & 0 & (110) & $Q_{xy}(0)$&\bf 0\\
4224.2& 4214.1	&$2.111 \pm 0.066$	& 1& (000) & 1 & (110) & $Q_{xy}(1)^c$&10.1\\
4227.2& 4214.1	&$10.143 \pm 0.067$	& 1& (000) & 1 & (110) & $Q_{xy}(1)^d$&13.1\\
%\hline
&&&&&&&&\\
4462.3& 4456.3	&$0.570 \pm 0.023$	& 0& (000) & 2 & (001) & $S_z(0)$	&6.0\\
4543.4& 4550.7	&$0.582 \pm 0.012$	& 0& (000) & 2 & (110) & $S_{xy}(0)$&-7.3\\
%\hline
&&&&&&&&\\
4678.8& 4667.0	&$0.831 \pm 0.012$	& 1& (000) & 3 & (001) & $S_z(1)$	&11.8\\
4761.3& 4761.3	&$1.900 \pm 0.019$	& 1& (000) & 3 & (110) & $S_{xy}(1)$&0.0\\
\end{tabular}
\end{ruledtabular}
\end{table*}
\endgroup

 Table\,\ref{tab:OPLowT_transitionsC70} summarizes the assignment of transitions observed in the experiment.
Since our model described in Theory section\,\ref{sec:seventy_ham} did not include translation-rotation coupling we used the transitions where $L$ and $J$ are both zero or one of them is zero in the initial and final states.
The fundamental frequency $\wzeroV=4069.1$\wn/ was chosen to match the  experimental fundamental \ortho/ transition $Q(1)$; the rotational constant $B_e=59.865$\wn/ and its corrections $\alpha_e=2.974$\wn/, $D_e=0.04832$\wn/ of \Htwoatsixty/  were used.
Next the two \para/ transitions were matched, $Q_z(0)$ and $Q_{xy}(0)$ to  obtain  $^v\!\wzeroTz=56.5$\wn/,  $^v\!\wzeroTxy=150.9$\wn/.
The 4220\wn/ line  was chosen as $Q_{xy}(0)$ because its $T$ dependence, Fig.\,\ref{fig:C70TdepIntensity}\,a), is similar to $S_z(0)$ and  $S_{xy}(0)$ line $T$ dependence.
The \ortho/ ground state splitting is 9.4\wn/ (7.4\wn/ from the 5D calculation \cite{Xu2009}) with $\vkappa=3.1$\wn/ which was obtained by
assuming that the $Q_z(1)^a$ transition,
$(000)E_1'\rightarrow (001)A_1'$ in Fig.\,\ref{fig:OPLowT_transC70},
is centered at 4110.3\wn/. 
It was assumed that the anharmonic corrections to translational energy and to $\vkappa$ are same for the $v=0$ and $v=1$ vibrational states.

The variation of the experimental frequencies compared to the model frequencies is within $\pm 10$\wn/, Table\,\ref{tab:OPLowT_transitionsC70}.
This is reasonable since our model did not include translation-rotation coupling and the  translation-rotation coupling in \Htwoatsixty/ creates splittings of this magnitude.
The vibrational frequency $\wzeroV$ is larger in \Htwoatseventy/ than in \Htwoatsixty/.
This difference must be taken with a reservation because the correction due to the difference of zero point translational energies in two vibrational states is not taken into account for \Htwoatseventy/ as it was done  for \Htwoatsixty/ \cite{MinGe2011}.
Again, a more elaborate model is needed for \Cseventy/ than the two oscillator model described in Theory section\,\ref{sec:seventy_ham}.

%==========================================
\section{Conclusions\label{sec:conclusions}}
%==========================================

IR absorption spectra of endohedral hydrogen isotopologs, \Htwo/ \cite{Mamone2009,MinGe2011},  \Dtwo/ \cite{Ge2011D2HD}, and HD \cite{Ge2011D2HD} in \Csixty/ are informative, involving excitations of vibrations, rotations and translational motion of dihydrogen.
The translational motion of the encapsulated molecule is quantized and coupled to its rotations because of the surrounding \Csixty/ cage.
The vibrational frequency of dihydrogen is redshifted compared to the gas phase value.
Together with the smaller rotational constant it shows that the hydrogen bond is stretched inside the cage and there is an attraction between H (or D) and C.
The heteronuclear HD does not rotate about its center of mass because of the surrounding cage. 
Different rotational and translational states are mixed and rotational quantum number $ J $ is not a good quantum number for \HDatsixty/.

Our study shows that  the vibrations and rotations of \Csixty/ and the crystal field effects of solid \Csixty/ are not important on the energy scale of   IR  measurements. 
If these effects are important their contribution to the IR spectra is the order of one wavenumber  splitting of few absorption lines.
Such small splitting is  consistent with the NMR \cite{Carravetta2006,Carravetta2007} and heat capacity \cite{Kohama2009} results. 

\Cseventy/ has ellipsoidal shape and this splits the translation-rotation states of \Htwo/.
The translational frequency is about 180\wn/ in \Htwoatsixty/.
In \Cseventy/ this three dimensional mode is split into a two dimensional mode at 151\wn/ and a one dimensional mode at 56\wn/.
The 5D quantum mechanical calculation \cite{Xu2009} is in very good agreement with this experimental result.

%=============================
\section{Acknowledgments}
%=============================

\begin{acknowledgments}
This research was supported by the Estonian Ministry of Education and Research grant SF0690029s09, Estonian Science Foundation grants  ETF8170, ETF8703, JD187.
\end{acknowledgments}

\bibliographystyle{apsrev}
%\bibliography{Fullerene,books}

\end{document}